\documentclass[reprint,showkeys,showpacs,superscriptaddress,nofootinbib,longbibliography,floatfix]{revtex4-2}
\usepackage[utf8]{inputenc}
\usepackage{textgreek}
\usepackage{comment}
\usepackage{lipsum}
\usepackage{natbib}
\usepackage{amsmath}
\usepackage{longtable}
\usepackage{orcidlink}
\usepackage{float}
\usepackage{rotating}
\usepackage{soul}
\usepackage{amssymb}
\usepackage{xcolor}
\usepackage{newunicodechar}
\newunicodechar{∼}{\sim}
\usepackage{enumerate}
\usepackage{array}
\usepackage{newunicodechar}
\newunicodechar{₃}{$_3$}
\usepackage{newunicodechar}
\newunicodechar{₂}{$_2$}
\usepackage{color}
\usepackage{xcolor}
\usepackage{tabu}
\usepackage{booktabs}
\usepackage{multirow}
\usepackage{graphicx}
\usepackage{simplewick}
\usepackage[dvipsnames]{xcolor} 
\usepackage{cancel}
\usepackage[margin=1in]{geometry}
\usepackage{bigints}
\usepackage{dirtytalk}
\usepackage{hyperref}
\definecolor{dark-red}{rgb}{0.9,0.15,0.15}
\definecolor{dark-blue}{rgb}{0.15,0.15,0.4}
\definecolor{medium-blue}{rgb}{0,0,0.5}
\hypersetup{
	colorlinks=true,
	citecolor=blue,
	linkcolor=red,
	filecolor=blue,      
	urlcolor=blue,
}
\begin{document}

\title{Interface and Strain \textcolor{black}{C}ontrol of \textcolor{black}{E}mergent Weyl Semimetallic Phase  in \texorpdfstring{SrNbO$_{3}$/LaFeO$_{3}$}{SrNbO3/LaFeO3} Heterostructures}

\author{Sairam Ithineni\,\orcidlink{0009-0006-6986-7317}}
\email{sairam.i@iopb.res.in}
\affiliation{Institute of Physics, Bhubaneswar-751005, India}
\affiliation{Homi Bhabha National Institute, Anushakti Nagar, Mumbai-400094, India}

\author{Pratik Sahu\,\orcidlink{0000-0003-0346-5682}}
\affiliation{Center for Atomistic Modelling and Materials Design, Indian Institute of Technology Madras, Chennai 600036, India}
\affiliation{Condensed Matter Theory and Computational Lab, Department of Physics, Indian Institute of Technology Madras, Chennai 600036, India}

\author{Soumyakanta Panda\,\orcidlink{0000-0001-6548-2010}}
\altaffiliation{Present address: Laboratoire Albert Fert, CNRS, Thales, Université Paris-Saclay, 91767 Palaiseau, France}
\affiliation{School of Basic Sciences, Indian Institute of Technology Bhubaneswar, Bhubaneswar, Odisha 752050, India}

\author{Aditya Mehta\,\orcidlink{0009-0003-2926-6793}}
\affiliation{Institute of Physics, Bhubaneswar-751005, India}
\affiliation{Homi Bhabha National Institute, Anushakti Nagar, Mumbai-400094, India}

\author{Debashree Nayak}
\affiliation{Homi Bhabha National Institute, Anushakti Nagar, Mumbai-400094, India}
\affiliation{School of Physical Sciences, National Institute of Science Education and Research, Jatni 752050, India}

\author{Amit Chauhan}
\affiliation{Peter Gr\"unberg Institute, Forschungszentrum J\"ulich, 52425 J\"ulich, Germany}

\author{Shwetha G Bhat\,\orcidlink{0000-0003-4237-9082}}
\affiliation{Department of Physics, Indian Institute of Science Bangalore-560012, India}

\author{Niharika Mohapatra\,\orcidlink{0000-0002-3182-8818}}
\affiliation{School of Basic Sciences, Indian Institute of Technology Bhubaneswar, Bhubaneswar, Odisha 752050, India}

\author{K. Senapati\,\orcidlink{0000-0003-1422-3195}}
\affiliation{Homi Bhabha National Institute, Anushakti Nagar, Mumbai-400094, India}
\affiliation{School of Physical Sciences, National Institute of Science Education and Research, Jatni 752050, India}

\author{B. R. K. Nanda\,\orcidlink{0000-0001-8330-4824}}
\affiliation{Center for Atomistic Modelling and Materials Design, Indian Institute of Technology Madras, Chennai 600036, India}
\affiliation{Condensed Matter Theory and Computational Lab, Department of Physics, Indian Institute of Technology Madras, Chennai 600036, India}

\author{D. Samal\,\orcidlink{0000-0003-2618-4445}}
\email{dsamal@iopb.res.in}
\affiliation{Institute of Physics, Bhubaneswar-751005, India}
\affiliation{Homi Bhabha National Institute, Anushakti Nagar, Mumbai-400094, India}

\date{\today}

\begin{abstract}
Realizing correlated topological semimetallic phases in bulk transition-metal oxides remains challenging due to rigid lattice symmetry, correlation-induced gap opening, and limited structural tunability. \textcolor{black}{However, complex-oxide thin films and heterostructures provide a powerful platform to stabilize topological phases by tailoring the requisite lattice symmetry through strain control and interface design}. In this study, we demonstrate the emergence of Weyl-like electronic states and associated chiral transport in SrNbO$_3$ (SNO)/LaFeO$_3$ (LFO) bilayers. Transport measurements reveal signatures consistent with nontrivial topology, including large non-saturating MR, a nonlinear Hall response, and a chiral anomaly like feature in longitudinal magnetotransport under parallel electric and magnetic fields ($\mathbf{B} \parallel \mathbf{I}$). In addition, we observe a \textcolor{black}{signature} of anomalous Hall contribution, likely arising from \textcolor{black}{proximity effect induced by LFO layers at the interface}. \textcolor{black}{First-principles calculations reveal an $a^0a^0c^-$ rotation pattern of the NbO$_6$ octahedra, together with interfacial lattice distortions in the SNO layer that drive the emergence of a twofold degenerate Weyl semimetallic phase protected by screw axis lattice symmetry.} \textcolor{black}{This is further confirmed by Berry curvature calculations, which show opposite sign Berry curvature peaks for the upper and lower band characteristic of a  Weyl node.} Our combined experimental and theoretical results highlight the critical role of strain and interfacial octahedral distortions in \textcolor{black}{stabilizing Weyl phase in transition metal based perovskite bilayer.}

\end{abstract}


\maketitle

\section{Introduction}
Dirac and Weyl semimetals constitute a compelling class of topological quantum materials that host symmetry or \textcolor{black}{topology protected} linearly dispersing degenerate band crossings near the Fermi energy (E$_{f}$)~\cite{yan2017topological,narang2021topology,wang2017quantum,hu2019transport,kumar2020topological}. The low energy \textcolor{black}{quasiparticles} in these topological \textcolor{black}{systems} are governed by relativistic Dirac or Weyl equations, rather than the conventional  parabolic  dispersion~\cite{armitage2018weyl,young2012dirac}. While Dirac semimetals host fourfold degenerate linear crossings stabilised by time-reversal, inversion, and crystal symmetries, Weyl semimetals exhibit twofold-degenerate linear band crossings under broken inversion and/or \textcolor{black}{time reversal} symmetry and always appear in pairs with opposite chirality acting as source and sink of Berry curvature in momentum space~\cite{lv2021experimental,wang2017quantum}. The distinctive band topology of these materials engenders unconventional transport responses, including ultrahigh carrier mobility~\cite{fujioka2019strong}, large nonsaturating magnetoresistance (MR)~\cite{chen2025extremely,singha2017large,leahy2018nonsaturating},  chiral anomaly~\cite{zyuzin2012topological, ong2021experimental,zyuzin2017magnetotransport}, Fermi-arc surface states~\cite{xu2015observation,wan2011topological}, and intrinsic anomalous Hall effect~\cite{ong2021experimental,takahashi2018anomalous,thakur2020intrinsic}. Following the discovery of three-dimensional Dirac semimetals in Na$_3$Bi~\cite{liu2014discovery, xiong2015evidence}and Cd$_3$As$_2$~\cite{neupane2014observation,li2016negative} and the subsequent realization of a Weyl semimetal phase in TaAs~\cite{lv2015experimental,yang2015weyl}, intense \textcolor{black}{efforts} have focused on identifying topological semimetals across a wide range of material systems~\cite{xu2015experimental,sun2015topological,souma2016direct,li2017evidence,deng2016experimental,zhang2019catalogue}. Most of the well established topological semimetals are weakly correleated, where $s$ and $p$ orbital derived bands near the Fermi energy contribute to electronic properties~\cite{liu2014discovery,samal2016molecular,nakamura2020robust,konig2007quantum,bernevig2006quantum,zhang2009topological}. Realizing topological semimetallic character in correlated $d$-electron based transition-metal oxides is particularly appealing, as the coexistence of nontrivial band topology and electron correlations in partially filled $d$ bands may \textcolor{black}{give rise to} emergent quantum phases such as Mott topological insulators~\cite{raghu2008topological}, axion insulators~\cite{liu2020robust}, and topological superconductors~\cite{mandal2023topological,huang2021review,fang2019discovery}, which are inaccessible in weakly interacting systems.\\
Oxide based correlated topological semimetallic states have been primarily reported in a limited class of iridate-based pyrochlore oxides, where the interplay of strong spin-orbit coupling of $5d$ electrons, sizable electron correlations, and geometrical frustration inherent to the lattice gives rise to nontrivial topology~\cite{yang2014emergent,ghosh2023spontaneous,yang2011quantum}. However, realizing such phases in a broader class of bulk transition-metal oxides is difficult, as \textcolor{black}{electronic correlations} and symmetry lowering due to lattice distortions often gap the system, rendering it topologically trivial. Notably, systems preserving non-symmorphic crystalline symmetries are robust against such perturbations and can protect nontrivial band topology~\cite{schoop2016dirac,li2018nonsymmorphic,meena2025nonsymmorphic,chen2016topological}. Therefore, \textcolor{black}{approaches} to enforce non-symmorphic symmetry is crucial for stabilizing topological phases. In this regard, thin films and heterostructures of complex oxides provide a distinct advantage, as epitaxial strain can be leveraged to tune lattice symmetry and structural distortions, enabling access to the symmetry conditions required to stabilise topological phases\textcolor{black}{,
 as demonstrated in model systems such as  CaIrO$_3$, SrIrO$_3$, SrRuO$_3$ and SrNbO$_3$.~\cite{masuko2019strain,liu2016strain, ok2021correlated,khetan2026stoichiometric,jaiswal2022direct,lin2021electric}} In particular, SNO films grown on SrTiO$_3$ (STO) along (001) direction host nontrivial band topology characteristic of a Dirac semimetallic state, driven by strain-induced non-symmorphic symmetry of the oxygen octahedral network~\cite{mohanta2021semi,ok2021correlated,rosendal2023octahedral,chikina2025intrinsic}, whereas bulk SNO behaves as a conventional metal. Angle-resolved photoemission spectroscopy measurements on SNO thin films reveal topologically protected Dirac band crossings near the Fermi level~\cite{chikina2025intrinsic}, resulting in large linear non-saturating magnetoresistance, ultrahigh carrier mobility ($\sim10^{4}$ cm$^{2}$\,V$^{-1}$\,s$^{-1}$), and a nonzero Berry phase. In contrast, SNO films grown on LSAT(001)~\cite{oka2015intrinsic} and KTO(001)~\cite{okuma2024large} exhibit conventional transport behavior, indicating that the topological states in SNO are highly sensitive to epitaxial strain induced octahedral rotations.\par
In this work, we demonstrate the emergence of a structurally tuned Weyl \textcolor{black}{semimetallic} state and chiral magnetotransport in SNO by interfacing it to an antiferromagnetic LaFeO$_3$ (LFO) in the form of SNO/LFO bilayers, where LFO serves as the underlayer. \textcolor{black}{LFO, a prototypical G-type canted antiferromagnetic insulator with a Néel temperature of $\sim 740$~K, crystallizes in the orthorhombic \textit{Pnma} structure and has a pseudo cubic lattice parameter \textcolor{black}{that closely matches with SrTiO$_3$}~\cite{coutinho2017structural}.} \textcolor{black}{Since both SNO and LFO crystallize in the orthorhombic structure, their structural  compatibility facilitates the epitaxial growth of one on the other.} The insulating nature of LFO ensures that charge transport is primarily confined to the SNO in  SNO/LFO bilayer. Structural characterization confirms epitaxial growth of the SNO/LFO heterostructure. Transport measurements reveal large non-saturating MR accompanied by high carrier mobility ($\sim10^{4}$ cm$^{2}$\,V$^{-1}$\,s$^{-1}$), indicative of topological electron transport. \textcolor{black}{Importantly, the SNO/LFO heterostructure exhibits a signature of the chiral anomaly characterized by  negative longitudinal MR ($\mathbf{B} \parallel \mathbf{I}$)}. First-principles calculations indicate an $a^{0}a^{0}c^{-}$ rotation pattern of the NbO$_6$ octahedra together with interfacial distortions \textcolor{black}{that stabilize the formation of Weyl nodes protected by screw axis symmetry}. The calculated band structure reveals twofold-degenerate states approximately 0.2~eV above the E$_{f}$ with opposite large Berry curvatures, consistent with the presence of  Weyl \textcolor{black}{node}.
  \begin{figure*}
    \centering
    \includegraphics[width=\linewidth]{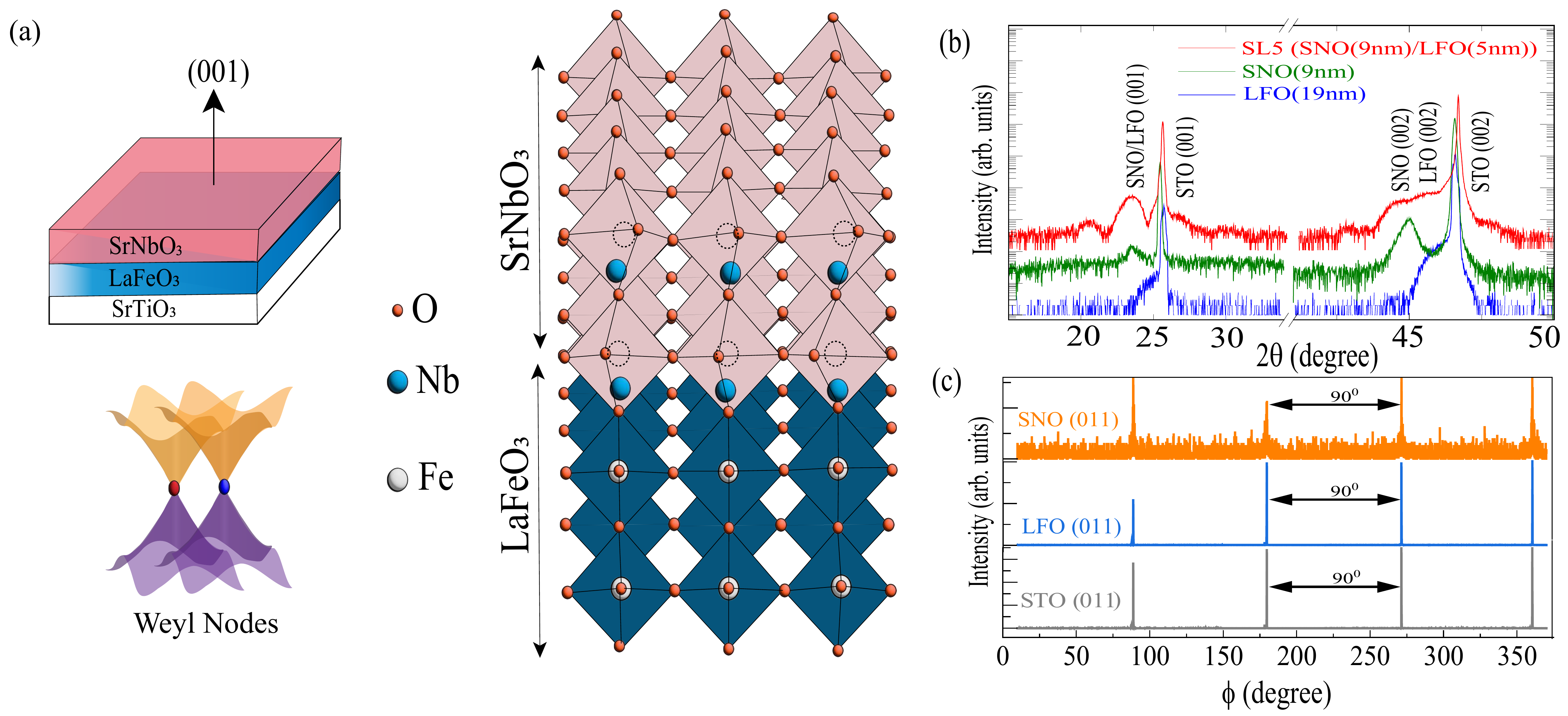}
   \caption{\textcolor{black}{(a) Top left schematic illustrates the SrNbO$_3$/LaFeO$_3$ bilayer grown on SrTiO$_3$(001). The right panel depicts the NbO$_6$ and FeO$_6$ octahedral networks in the bilayer, with Sr and La ions excluded for visual clarity. Epitaxial strain induces $a^{0}a^{0}c^{-}$ type octahedral rotations, while interfacial octahedral distortions lead to Nb off-center displacements, breaking inversion symmetry and generating Weyl nodes near the Fermi level in SNO, as shown in the bottom left figure.} (b) 2$\theta$–$\theta$ \textcolor{black}{XRD pattern} of the SL5 bilayer and \textcolor{black}{bare SNO and LFO layers} grown on STO(001), showing the (00l) reflections. (c) $\phi$-scan measurements \textcolor{black}{on} the (011) reflections for the SNO and LFO layers and the STO substrate, demonstrating the expected fourfold rotational symmetry and epitaxial in-plane alignment.}
    \label{fig:1}
\end{figure*}
\section{Experimental Section}
\subsection {Sample Growth:}
Thin films of  SNO(9 nm)/LFO(5 nm), SNO(9 nm)/LFO(22 nm) bilayers \textcolor{black}{as well as, bare SNO and LFO layers} were grown on (00$1$) oriented STO substrates using PLD, with a KrF excimer laser ($\lambda = 248~nm$), operating at a repetition rate of 4 Hz and a laser energy of 400~mJ. \textcolor{black}{Sr$_{2}$Nb$_{2}$O$_{7}$ and LaFeO$_{3}$ targets were used for the growth of the respective layers}. The substrate temperature was maintained at 650$^{0}$~C for all depositions. SNO films were grown under a base pressure of 2$\times 10^{-6}$~mbar. In the case of the SNO/LFO bilayer, the LFO layer was initially deposited at 650$^{0}$~C, under an oxygen pressure of  2$\times 10^{-3}$~mbar, followed by deposition of the SNO layer at the same temperature under a base pressure of 2$\times 10^{-6}$~mbar.~ For the sake of brevity, SNO(9 nm)/LFO(5 nm), SNO(9 nm)/LFO(22 nm) are referred to as SL5 and SL22, respectively.
\subsection{X-ray and Transport Characterization :}
The structural characterization of the films were examined using Rigaku SmartLab High-resolution X-ray diffraction (HR-XRD) with Cu~Kα$_{1}$ radiation source ($\lambda = 1.54$~\AA).  Thin films were wire-bonded using Al wires \textcolor{black}{for} electrical transport measurements.~\textcolor{black}{Magnetotransport measurements were performed in a \textcolor{black}{CRYOGENIC} physical
property measurement system (PPMS). The magnetic field was applied both parallel and
perpendicular to the film surface.}\\
\subsection{Method of DFT calculations}
The theoretical calculations in this work were performed using the following approach. In the case of structural relaxation calculation for SNO/LFO heterostructure, density functional theory calculations were performed using the Vienna Ab initio Simulation Package (VASP)~\cite{kresse1993ab,kresse1996efficient},  employing the generalized gradient approximation in the form of the Perdew-Burke-Ernzerhof (PBE) functional~\cite{perdew1996generalized}. The structures were relaxed until the residual forces on each atom were less than $10^{-4}$~eV/\AA. On the other hand, the topological properties were evaluated using tight-binding models constructed from hopping parameters obtained from maximally localized Wannier functions generated with Wannier90~\cite{mostofi2008wannier90}. The Wannier functions were converged with a localization tolerance of $10^{-10}$~\AA. The underlying self-consistent field calculations were performed using Quantum ESPRESSO~\cite{giannozzi2009quantum} with a plane-wave energy cutoff of 200~Ry and an electronic energy convergence threshold of $10^{-9}$~eV.

\section{Results and Discussion} 
\subsection{Structural Characterization :} 
\textcolor{black}{Thin films of SNO/LFO bilayers SNO(9~nm)/LFO(5~nm), SNO(9~nm)/LFO(22 nm) along with bare SNO and LFO layers, were epitaxially grown on (001) oriented STO substrates using pulsed laser deposition (PLD).~ For the sake of brevity, SNO(9 nm)/LFO(5 nm), SNO(9 nm)/LFO(22 nm) are referred to as SL5 and SL22, respectively.} 
\textcolor{black}{SNO/LFO bilayer heterostructures were epitaxially grown on STO substrates along (001) direction.\textcolor{black}{Bulk SNO crystallizes in an orthorhombic structure (space group $Pnma$) with lattice parameters $a = 5.6894~\text{\AA}$, $b = 5.6944~\text{\AA}$, and $c = 8.0684~\text{\AA}$~\cite{asmara2017tunable,ok2021correlated}, corresponding to a pseudocubic lattice parameter $a_{\mathrm{pc}} \approx 4.02~\text{\AA}$. \textcolor{black}{S}imilarly LFO adopts an orthorhombic $Pnma$ structure with lattice constants $a = 5.57~\text{\AA}$, $b = 5.565~\text{\AA}$, and $c = 7.845~\text{\AA}$~\cite{coutinho2017structural}, corresponding to a pseudocubic lattice parameter of $a_{\mathrm{pc}} \approx 3.93~\text{\AA}$.~\cite{chen2013significant,panda2025tuning}} \textcolor{black}{When grown on STO(001), which has a lattice parameter of $3.905~\text{\AA}$, SNO experiences a compressive lattice mismatch strain of approximately $3.02\%$, whereas LFO exhibits a significantly smaller lattice mismatch of about $0.64\%$. \textcolor{black}{Such epitaxial strain modifies the B–O bond length and the B–O–B bond angle in the ABO$_3$ perovskite oxide structure from their equilibrium values, thereby inducing structural distortion}. In the SNO/STO system, the compressive epitaxial strain imposed by the substrate has been shown to modify the NbO$_6$ octahedral network through $a^{0}a^{0}c^{-}$/ $a^{0}a^{0}c^{+}$ type octahedral rotations}, \textcolor{black}{which give rise to Dirac crossing near $E_{f}$}~\cite{mohanta2021semi,ok2021correlated,rosendal2023octahedral,chikina2025intrinsic}. \textcolor{black}{However, in case of  \textcolor{black}{present study on} SNO/LFO/STO heterostructure, the LFO buffer layer introduces distinct structural distortions in the SNO layer, as evidenced by First-principles calculations discussed in a later section. It is found that the NbO$_6$ octahedra undergo an out-of-phase $a^{0}a^{0}c^{-}$ type rotation, in addition to interfacial distortion where the Nb atom is displaced downward.\hspace{-0.02em} Further, the interfacial structure retains nonsymmorphic symmetry with a screw axis along the $b$ direction. This lowering of local symmetry in the SNO/LFO bilayer leads to the emergence of Weyl nodes at the interface, as illustrated schematically in Fig.~1(a).}}\par
\textcolor{black}{In order to check the epitaxial and oriented growth of LFO and SNO layers, we have performed \textcolor{black}{X-ray diffraction (XRD)} measurements.   X-ray reflectivity measurements were performed to assess the film thickness, as shown in Fig.~S2~\cite{supple}. The 2$\theta$–$\theta$ scan shown in Fig.~1(b) for the SL5 bilayer presents the $(00l)$ peaks of both the SNO and LFO layers, indicating $c$-axis oriented growth of the films on the STO (001) substrate. \textcolor{black}{The same is also shown for bare SNO (9~nm) and LFO (19~nm) films for reference.} The out-of-plane \textcolor{black}{c-axis} lattice parameters extracted from the 2$\theta$–$\theta$ scan of SL5 are 4.03~nm and 4.10~nm for LFO and SNO, respectively. \textcolor{black}{The c-axis expansion \textcolor{black}{is expected to} result from the BO$_{6}$ octahedra distortions induced by the in-plane compressive strain in the respective layers}. Additionally, the epitaxial growth of the SNO and LFO layer in the SNO/LFO/STO heterostructure was verified by performing phi-scans around the (011) reflection (Fig.~1(c)). A clear fourfold rotational symmetry of (011) reflections consistent with the symmetry of the substrate is observed, indicating a cube-on-cube epitaxial relationship between the heterostructure and STO. The transport properties of the heterostructures are discussed next, focusing primarily on SL5, while SL22 exhibits qualitatively similar behavior, as shown in Fig.~S3~\cite{supple}.}
\textcolor{black}{\subsection{Electron Transport}}  
Fig.~\ref{fig:2}(a) presents the temperature-dependent resistivity in the range of 2–60 K \textcolor{black}{of sample SL5}, and its variation is described by $\rho(T) = \rho_{0} + A T^{n}$ with $n \approx 2$, where $\rho_{0}$ represent the residual resistivity and the coefficient $A$ is a constant associated with quasiparticle scattering rate. 
The value of the residual resistivity $\rho_{0}$ and coefficient $A$ extracted from the fit for the bilayer are $0.102 ~\mu\Omega \,\text{cm}$ and $5.54 \times 10^{-6}~\mu\Omega \,\text{cm K}^{-2}$ respectively. These values are consistent with earlier findings for SNO.~\cite{oka2015intrinsic}

\textcolor{black}{The inset of Fig.~\ref{fig:2}(a) shows the temperature and field dependent longitudinal resistivity $\rho_{xx}(T,\mu_{0}{H})$  under the \textcolor{black}{field applied}  along the out-of-plane direction of the film.} For magnetic fields above 1~T, the longitudinal resistivity $\rho_{xx}(T,\mu_{0}H>$1~T) develops an upturn as the temperature is lowered below 25~K, followed by a reentrant metallic behavior at lower temperatures. To investigate the origin of \textcolor{black}{upturn} behavior, we \textcolor{black}{invoke} a semiclassical \textcolor{black}{approach}, within which the field-dependent resistivity is described by the \textcolor{black}{relation~\cite{PhysRevB.92.180402}}
\begin{equation}
\rho_{xx}(T,\mu_{0}H) = \rho_{xx}(T,0) + \frac{\alpha H^m}{\rho_{xx}(T,0)^{m-1}}.
\end{equation}
where, $\rho_{xx}(T,0)$ denotes the temperature-dependent zero-field resistivity, which is inversely proportional to the charge carrier mobilities $\mu_{e,h}(T)$ \textcolor{black}{($\rho_{xx}(T,0) = 1/({e\left(n_e \mu_e + n_h \mu_h\right)})$)}, \textcolor{black}{m is the power-law exponent, $\alpha$ is the MR scaling factor}~\cite{PhysRevB.92.180402}. In the present case, the charge carrier mobilities is relatively high at low temperatures ($10^{4}~\text{cm}^{2} \text{V}^{-1}\text{s}^{-1}$) as discussed later in this section. This results in a substantial reduction in $\rho_{xx}(T,0)$, making the first term negligible. Consequently, the field-dependent term becomes dominant, leading to an upturn in the resistivity at low temperatures, a behavior commonly observed in high mobility semimetals, both topologically trivial and nontrivial semimetals~\cite{PhysRevB.92.180402, tafti2016resistivity,  PhysRevLett.115.046602}. At lower temperatures below 10~K, the resistivity decreases again, signaling a reentrant metallic behavior, similar to that reported in the Weyl semimetals NbP and TaP~\cite{PhysRevB.96.115152,doi:10.1073/pnas.1808747115}.
Recently, Zhang et al~\cite{zhang2025inadequacy} proposed a unified framework to account for such reentrant behavior by considering temperature  and field \textcolor{black}{evolution} of Fermi surface topology that govern the underlying scattering rate and  shape the $\rho_{xx}(T)$ curve in magnetic field.
\begin{figure*}
    \centering
    \includegraphics[width=\linewidth]{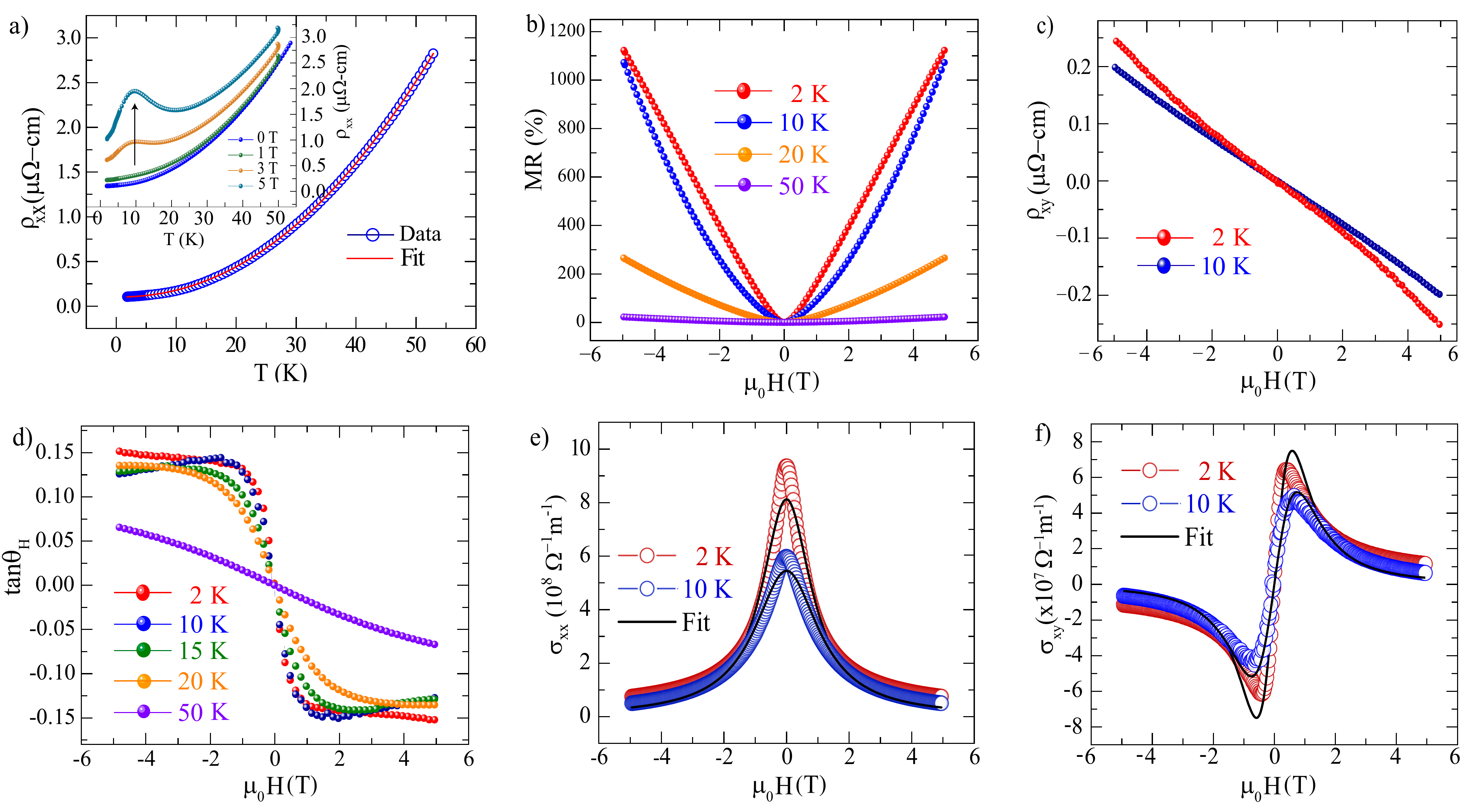}
   \caption{(a) Temperature dependence of electrical resistivity $(\rho_{xx})$ for SL5. The solid line represents a fit to the generalised power-law form \textcolor{black}{$\rho(T) = \rho_{0} + A T^{n}$}. The upper inset shows the temperature-dependent resistivity $\rho_{xx}$ of the sample measured under magnetic fields ranging from 0 to 5 T. (b) Field-dependent longitudinal resistivity $\rho_{xx}$ at \textcolor{black}{different temperatures (2~K,10~K, 20~K and 50~K)} as a function of the applied magnetic field. (c) Field-dependent Hall resistivity \textcolor{black}{($\rho_{xy}$)} measured at 2 K and 10 K as a function of the applied magnetic field. (d) Hall angle as a function of the applied magnetic field at various temperatures. (e) Field-dependent longitudinal conductivity $\sigma_{xx}$(H) at 2~K and 10~K. (f) Field-dependent Hall conductivity $\sigma_{xy}$(H) at 2~K and 10~K.}
    \label{fig:2}
    \label{fig:2}
\end{figure*}

\begin{figure*}
    \centering
    \includegraphics[width=\linewidth]{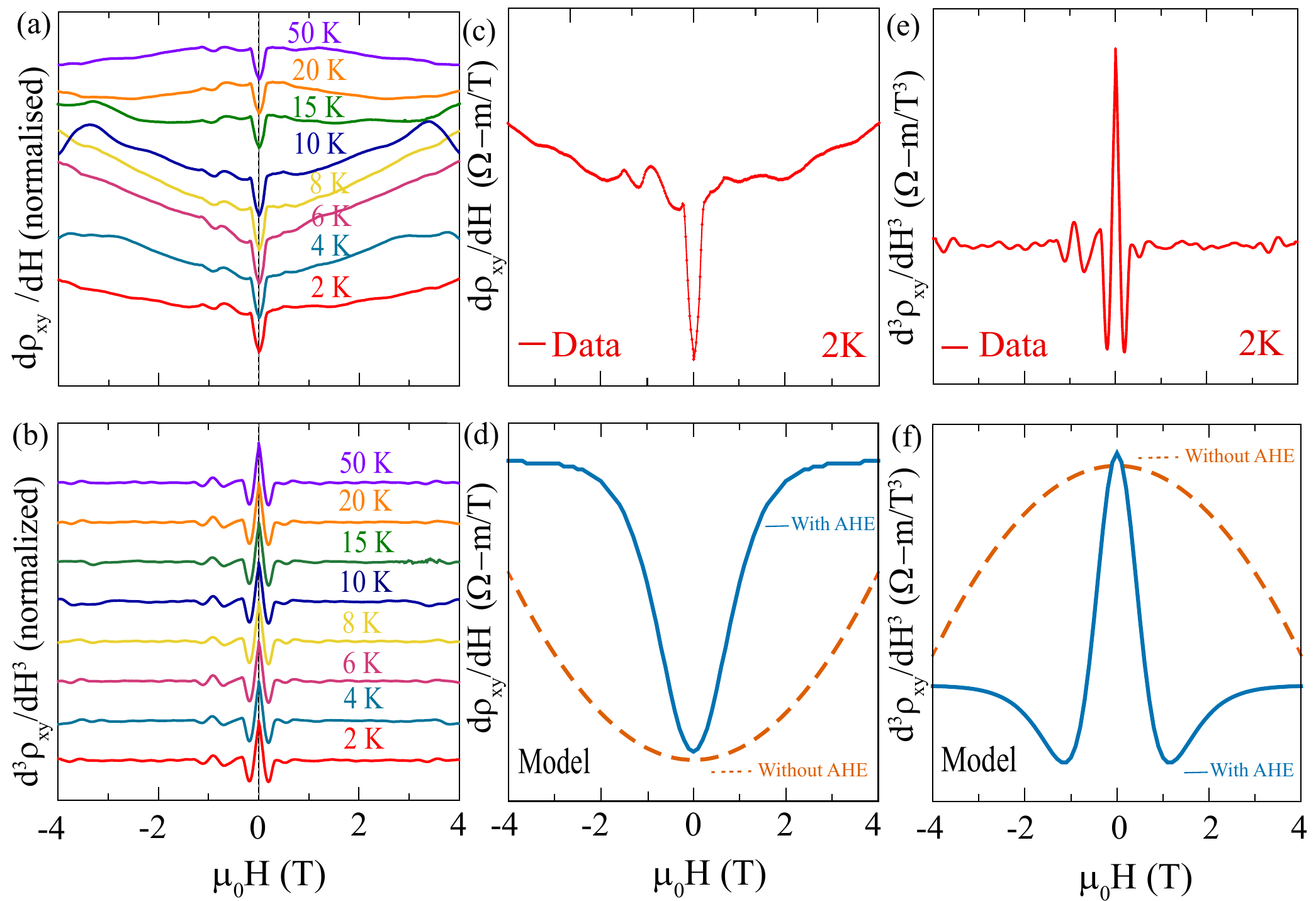}
   \caption{
(a) First derivative $d\rho_{xy}/dH$ highlighting a distinct dip centered at zero field.
(b) Third derivative $d^{3}\rho_{xy}/dH^{3}$, emphasizing the sharp zero-field structure across different temperatures.
(c) and (e) Representative $d\rho_{xy}/dH$  and $d^{3}\rho_{xy}/dH^{3}$, curves at 2~K for \textcolor{black}{SL5}. 
(d) and (f) Simulated $d\rho_{xy}/dH$ without (dashed) and with (solid) an anomalous Hall contribution, demonstrating how a finite anamolous component leads to a characteristic dip around zero field. }
    \label{fig:3}
\end{figure*}

Fig.~\ref{fig:2}(b) displays the field dependence of the longitudinal MR measured at various temperatures, with the magnetic field applied along the out-of-plane direction. At 2~K, the system exhibits a large positive MR  that increases almost linearly with magnetic field up to the highest applied field, reaching about 1200 \% at 5 T without any sign of saturation. Such a large, non-saturating linear MR is commonly attributed to two distinct mechanisms: quantum linear MR~\cite{abrikosov1998quantum,hu2008classical} and guiding-center model~\cite{song2015linear}. In Dirac semimetals, quantum linear MR arises when all charge carriers occupy the lowest Landau level \textcolor{black}{(quantum limit)}, and typically requires ultralow carrier concentrations on the order of $10^{17}$~cm$^{-3}$~\cite{lonchakov2024quantum,ok2021correlated}. In contrast, the guiding-center model attributes linear MR to the drift of charge carriers along curved trajectories in an inhomogeneous potential landscape and predicts a linear field dependence when the Hall angle saturates at low temperatures~\cite{song2015linear,christensen2024extreme,zhang2021extremely}. \textcolor{black}{For SL5, the carrier concentration is $\sim10^{21}$~cm$^{-3}$, as estimated from the conductivity analysis (shown  in Fig.~\ref{fig:2}(e) and Fig.~\ref{fig:2}(f))}, \textcolor{black}{and it} is far too large to satisfy the quantum-limit condition. Moreover, the low-temperature saturation of the Hall angle observed in Fig.~\ref{fig:2}(d) is characteristic of the guiding-center regime and is consistent with earlier reports on SNO.~\cite{zhang2021extremely} Upon increasing the temperature to 10 K, the MR shows a slight deviation from linearity while maintaining a large magnitude of approximately 1100 \% at 5 T. As the temperature increases further to 50~K, the overall MR magnitude decreases progressively, and the field dependence changes from linear to \textcolor{black}{parabolic.}
The Hall resistivity shown in Fig.~\ref{fig:2}(c), displays a pronounced nonlinearity at low temperatures, indicative of multiband transport. Within the semiclassical description, the  conductivity tensor in it's complex form can be expressed as:
\begin{equation}
    \sigma = \frac{e n_e \mu_e}{1 + i \mu_e \mu_0 H} + \frac{e n_h \mu_h}{1 - i \mu_h \mu_0 H},
\end{equation}
\textcolor{black}{Where $n_{e}$ and $n_{h}$ represent the carrier concentrations, and $\mu_{e}$ and $\mu_{h}$ represent the mobilities of electrons and holes, respectively. 
The real and imaginary components of this
tensor correspond to the longitudinal and transverse conductivities, respectively, and are given by
\begin{equation}
    \sigma_{xy} =   \frac{eH\mu_0n_h \mu_h^2}{1 + \mu_h^2 \mu_0^2 H^2} -  \frac{n_ee \mu_0 H \mu_e^2}{1 + \mu_e^2 \mu_0^2 H^2},
\end{equation}
\begin{equation}
    \sigma_{xx} =  \frac{e n_h \mu_h}{1 + \mu_h^2 \mu_0^2 H^2} + \frac{e n_e \mu_e }{1 + \mu_e^2 \mu_0^2 H^2}.
\end{equation}
The conductivities are obtained from the measured resistivities $\rho_{xx}$ and $\rho_{xy}$ 
using $\sigma_{xy} = \frac{-\rho_{xy}}{\rho_{xx}^2+\rho_{xy}^2}$ and  $\sigma_{xx} = \frac{\rho_{xx}}{\rho_{xx}^2+\rho_{xy}^2}$. Both $\sigma_{xy}(H)$ and $\sigma_{xx}(H)$ are \textcolor{black}{simultaneously fitted to estimate $n_{e}$, $n_{h}$, $\mu_{e}$ and $\mu_{h}$} using a common set of parameters, as shown in \ref{fig:2}(e,f).}
\textcolor{black}{At 2~K, SL5 exhibits carrier concentrations ($n_e$ and $n_h$ $\sim$ $10^{21}$~cm$^{-3}$) and mobilities ($\mu_e$ and $\mu_h$ $\sim$ $10^{4}$~cm$^{2}$$\text{V}^{-1}\text{s}^{-1}$), and their variation with temperature upto 20~K is shown in Fig.~S4~\cite{supple}. Broadly, the extracted parameters are consistent with earlier reports on }\textcolor{black}{SNO.~\cite{ok2021correlated,zhang2021extremely}} However, a noticeable deviation of the fit from the two-band model is observed, which likely originates from an anomalous Hall contribution due to interfacial magnetism induced in SNO via proximity to the  LFO layer. Although a distinct AHE is not directly visible due to the dominant nonlinear Hall contribution from two-band transport, a derivative-based approach has been undertaken to delve into the underlying behavior~\cite{zhang2017magnetic,lazrak2025spontaneous}.

Fig.~\ref{fig:3}(a) displays the first derivative of the Hall resistivity, $d\rho_{xy}/dH$, which exhibits a clear dip centered at zero field across all measured temperatures. To further highlight this zero-field \textcolor{black}{feature}, the third derivative of Hall response has been investigated, and is plotted in Fig.~\ref{fig:3}(b). The third derivative enhances the inflection characteristics, revealing a distinct dip–peak-dip structure around zero field that remains visible across all the temperatures. This zero-field feature could possibly indicate an anomalous Hall contribution originating from the interfacial ferromagnetic component. To further examine the origin of the features observed in the Hall response, we simulated the Hall resistivity at \textcolor{black}{2~K} using a standard two-band model. 
\[
\rho_{xy} = 
\frac{
(\mu_{e}^{2} n_{h} - \mu_{h}^{2} n_{e}) 
+ (\mu_{e} \mu_{h} )H^{2} (n_{h} - n_{e})
}{
e \left( \mu_{e} |n_{e}| + \mu_{h} |n_{h}| \right)^{2}
+ (\mu_{e} \mu_{h} H)^{2} (n_{h} - n_{e})^{2}
} \, H
\]

\begin{figure*}
    \centering
    \includegraphics[width=\linewidth]{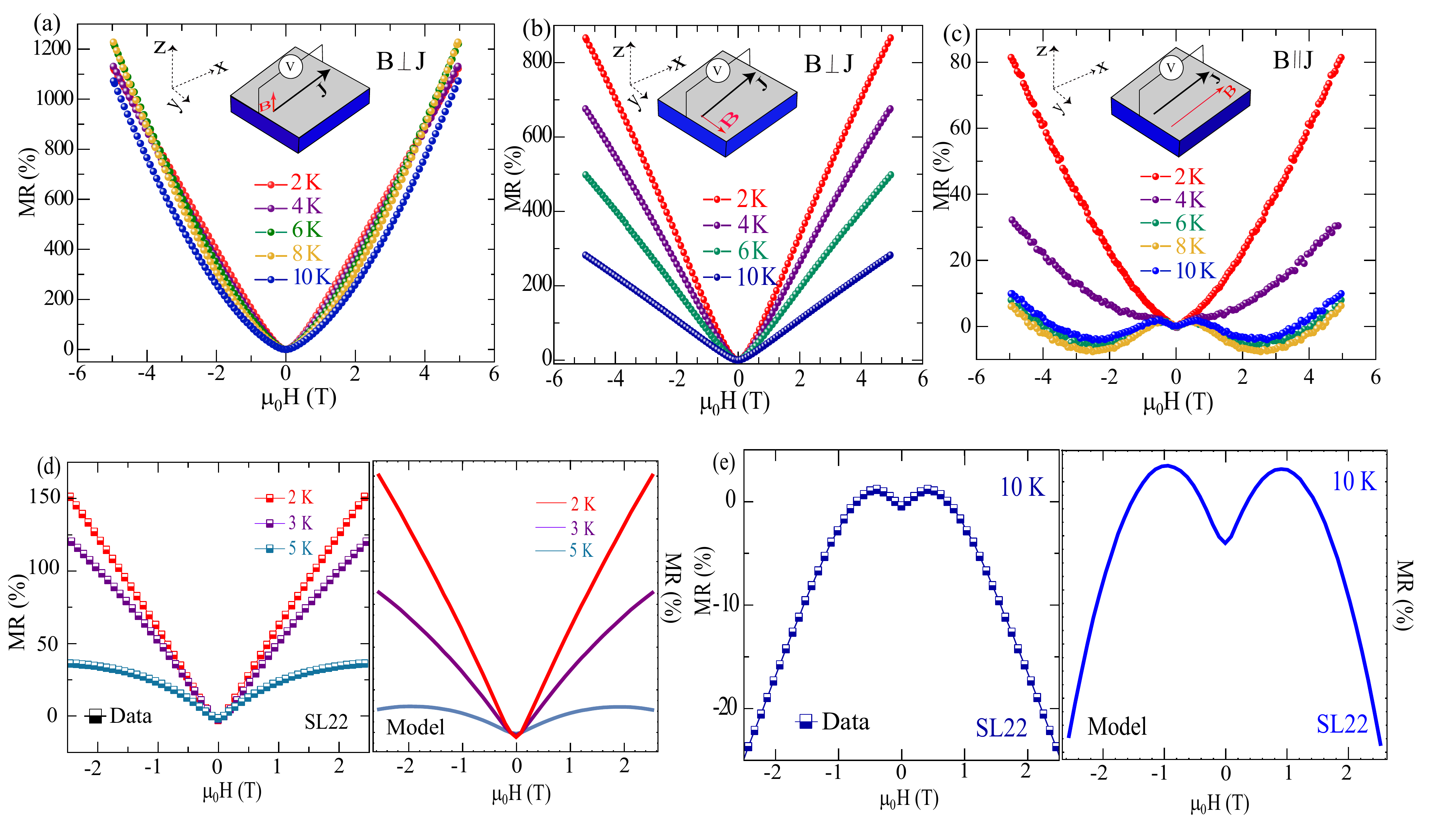}
  \caption{MR of the \textcolor{black}{SL5} bilayer  measured under different magnetic field orientations. 
(a) MR for \(\mathbf{B} \perp \mathbf{J}\) with \(\mathbf{B}\) applied along the out-of-plane direction in \textcolor{black}{bilayer SL5}. 
(b) MR for \(\mathbf{B} \perp \mathbf{J}\) with \(\mathbf{B}\) applied along the in-plane direction of \textcolor{black}{SL5}. 
(c) MR for \(\mathbf{B} \parallel \mathbf{J}\) with current and field along the in-plane direction of \textcolor{black}{SL5}. 
(d )MR of the SL22 bilayer for \(\mathbf{B} \parallel \mathbf{J}\)
 with current and field applied along the in-plane direction at 2, 3, and 5 K and  Modeled MR at 2, 3, and 5 K obtained from Mathematica. (e) Experimental MR at 10 K, along with Modeled MR at 10 K.}
    \label{fig:4}
\end{figure*}
As shown in Fig.~\ref{fig:3}(d), the first derivative of the Hall signal obtained from the two-band model without an anomalous Hall contribution (red dashed line) exhibits a broad and smooth field dependence. In contrast, when an additional anomalous Hall term of the form $-A \tanh(H/H_{0})$ (where H$_{0}$ sets the characteristic field scale over which the AHE like contribution saturates) is included to mimic the effect of interfacial magnetism in SNO, the simulated first derivative develops a sharp dip centered at zero field (blue solid line). A similar enhancement is also evident in the third derivative, as shown in \textcolor{black}{Fig.~\ref{fig:4}(f)}. These simulations \textcolor{black}{reflects} the characteristic dip–peak-dip structure observed experimentally in the derivative of the Hall resistivity, indicating that the nonlinear features in the derivative data are consistent with the presence of an additional Hall contribution beyond the conventional two-band response~\cite{zhang2017magnetic,lazrak2025spontaneous}. However, the persistence of the magnetic ordering beyond room temperature in LFO makes it challenging to disentangle the anomalous Hall contribution from the intrinsic nonlinear Hall response. As shown in Fig.~S5, the M-H isotherm ($M$–$H$) of the SL5 bilayer exhibits hysteretic magnetic response upto 300~K~\cite{supple}. The overall \textcolor{black}{Hall} transport behaviour indicates that the bilayer exhibits multiband semimetallic conduction, with an additional \textcolor{black}{anomalous Hall component arising from proximity-induced interfacial magnetism at the SNO/LFO interface, \textcolor{black}{which} is discussed in theory section.} \newline
\subsection{\textcolor{black}{Chiral Anomaly Induced Transport}} 
\vspace{-0.5em}
A Dirac node is a fourfold-degenerate linear band crossing that can be viewed as the superposition of two Weyl nodes with opposite chirality. Upon breaking either time reversal symmetry or inversion symmetry, the Dirac node splits into a pair of Weyl nodes, each carrying a definite chirality ($\chi = \pm 1$)~\cite{armitage2018weyl}. In the presence of a strong magnetic field, states near each Weyl node are quantized into Landau levels, with the lowest Landau level (LLL) being chiral and dispersing strictly parallel or antiparallel to the magnetic field, depending on the chirality $\chi$ of the Weyl node~\cite{armitage2018weyl,ong2021experimental}. Classically, the number of fermions of each chirality is conserved, and hence no net charge transfer occurs between Weyl nodes. However, when electric and magnetic fields are applied in parallel ($\mathbf{E} \parallel \mathbf{B}$), charge is pumped between Weyl nodes of opposite chirality given by ~\cite{ong2021experimental}:
\begin{equation}
 \frac{d n_{R/L}}{dt} = \pm \frac{e^2}{h^2}\, \mathbf{E}\cdot\mathbf{B}   
\end{equation}
leading to the violation of chiral charge conservation and the generation of an \textcolor{black}{axial current}. This manifests experimentally as a negative longitudinal MR, which is widely recognized as a hallmark of the chiral anomaly  \textcolor{black}{related transport in topological semimetals}~\cite{li2016chiral,li2016negative,ong2021experimental}.\\
Fig.~\ref{fig:4}(a)–(c) shows the magnetic field-dependent MR measured along three different configurations: (a) with current along the in-plane direction and magnetic field applied out-of-plane (out-of-plane MR), (b) with the in-plane magnetic field perpendicular to the current (inplane-transverse MR), and (c) with the in-plane magnetic field parallel to the current (inplane-longitudinal MR). In both the configurations, \textcolor{black}{i.e., Fig.~\ref{fig:4}(a) and Fig.~\ref{fig:4}(b)}, we observe a large unsaturating linear MR, reaching values as high as 1200\%. Interestingly, the in-plane longitudinal MR  ($\mathbf{E} \parallel \mathbf{B}$) in configuration Fig.~\ref{fig:4}(c) displays a different evolution from that of configuration in Fig.~\ref{fig:4}(a) and Fig.~\ref{fig:4}(b). While the in-plane longitudinal MR at 2~K remain positive, at higher temperatures the MR initially decreases with field and then exhibits a smooth upturn, beyond a characteristic field. Such behavior, featuring a partial negative MR component, has been reported in several topological systems and is frequently associated with the chiral anomaly~\cite{xiong2015evidence,li2016chiral,li2015giant}.
To further ascertain the presence of the chiral anomaly and confirm its reproducibility, we examined SL22 \textcolor{black}{(SNO(9~nm)/LFO (22~nm))}, whose transport characteristics closely resemble those of SL5 (as shown in Fig.~S4~\cite{supple}). The left hand plots in Fig.~\ref{fig:4}(d) and  Fig.~\ref{fig:4}(e) shows the in-plane logitudinal MR ($\mathbf{E} \parallel \mathbf{B}$) for SL22 at different temperatures and a clear indications of chiral anamoly like response is observed at 10~K.
To model the observed in-plane crossover, we employed a qualitative phenomenological form given by:
\[ 
MR_{\mathrm{total}} = \alpha * MR_{\mathrm{pos}} + (1-\alpha) * MR_{\mathrm{neg}} 
\]
where $\alpha$ represents the weighting factor associated with the positive MR component. In this framework, the negative contribution is modeled  as a $-B^{2}$ term arising from the chiral anomaly.~\cite{huang2015observation} Fig.~\ref{fig:4}(d) and Fig.~\ref{fig:4}(e) demonstrate that the proposed model qualitatively captures the experimental MR behavior of the SL22 bilayer at 2, 3, 5, and 10\,K. At 10\,K, the system predominantly exhibits a negative MR associated with the chiral anomaly. However, upon lowering the temperature, a competing positive MR component gradually emerges and, by 2\,K, completely outweighs the negative contribution. Although the chiral anomaly driven negative MR is typically expected to become more pronounced at low temperatures, the positive MR component \textcolor{black}{possibily arising from bulk SNO layer} grows more rapidly \textcolor{black}{that} suppresses the negative component.
The robust negative MR observed at 10\,K for in-plane longitudinal MR, and its absence for in-plane transverse MR, provides compelling evidence for the presence of the chiral anomaly in the SNO/LFO bilayer. \textcolor{black}{Interestingly, no signature of chiral anomaly induced negative MR is observed in the bare SNO film grown under identical conditions (Fig.~S7~\cite{supple}), rather it shows only large positive MR at low temperatures. This highlights the critical role of electronic structure at SNO/LFO interface in controlling the transport properties.}
\textcolor{black}{To examine the presence of non trivial band topology in the SNO/LFO bilayer, we  carried out first principle calculations, as discussed in next section.}
\subsection{First Principles calculations}
In this section, we theoretically investigate the electronic structure and topological properties of the SNO/LFO heterostructure using density functional theory (DFT) and Wannier based tight-binding models. \textcolor{black}{We construct a heterostructure with 4 layers of LFO in orthorhombic phase with lattice constants $a=5.56$~\AA, $b=5.67$~\AA, and $c=7.93$~\AA, and cubic SNO with lattice constant $a=4.02$~\AA.} 
Each LFO layer is made up of La$_{4}$Fe$_{4}$O$_{16}$, and each SNO layer consists of Sr$_{2}$Nb$_{2}$O$_{6}$ as formula unit cells. The Fe atoms are arranged in a G-AFM order. 
\begin{figure*}[!htb]
    \centering
    \includegraphics[width=\linewidth]{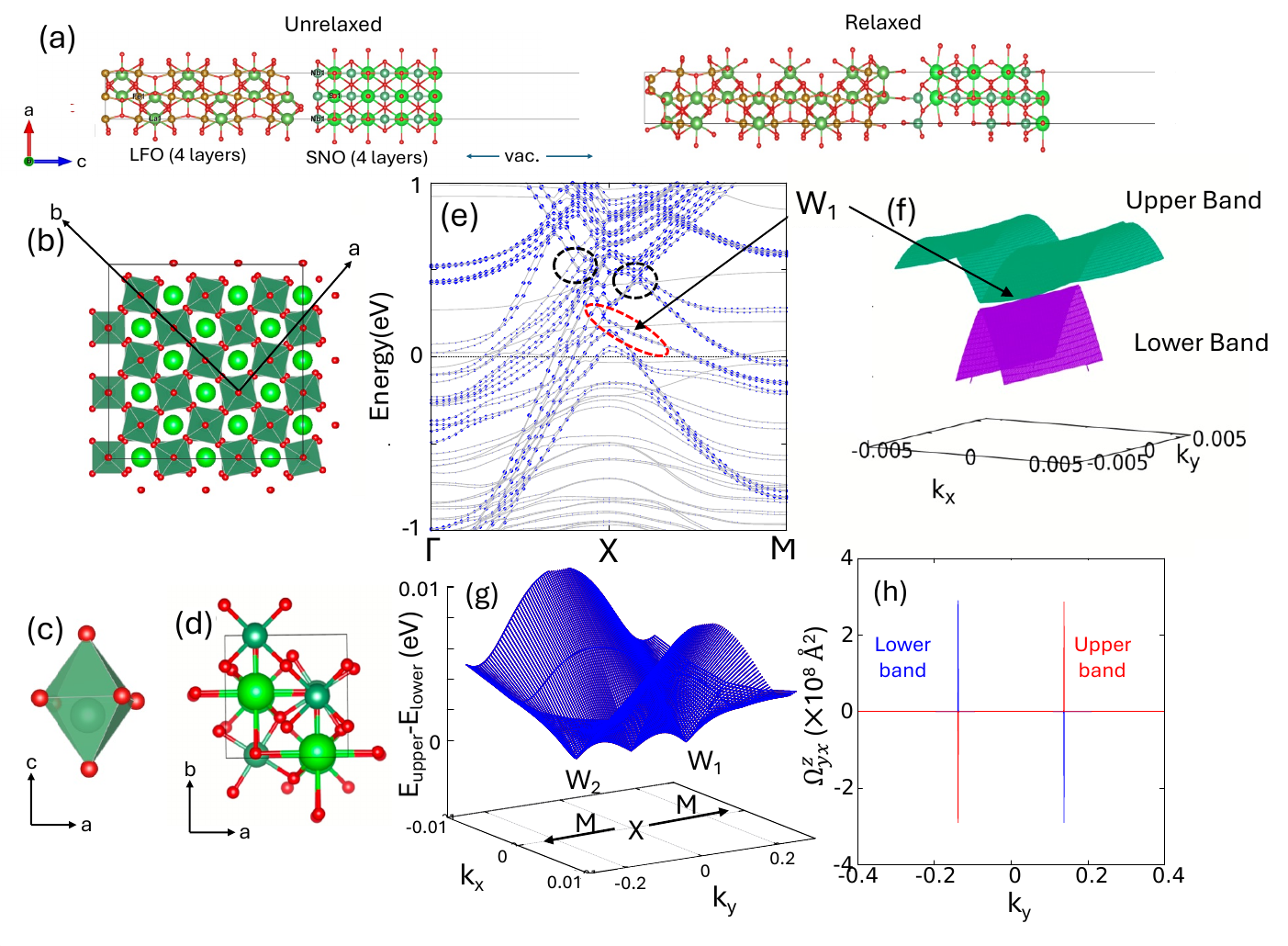}
     \caption{(a) LFO-SNO heterostructures (4 layers each) before and after relaxation. (b) Out of phase octahedral rotation ($a^0a^0c^-$) of the interfacial SNO layers. (c) The distorted SNO octahedra at the interface for the relaxed system, with the central Nb atom shifted downwards by 0.46~\AA~ having (d) screw axis symmetry along $\vec{b}$. (e) Band structure of the relaxed 4-layers SNO/LFO heterostructure, where the blue dots indicate the SNO characters, and linear band crossings are highlighted as dashed circles. From the layer projected band structure (as shown in Fig.~S9~\cite{supple}), the crossing $W_1$ is formed by the bands emerging from the interfacial layer (f). 3D plot of the band structure for the bulk system replicating the distortion occurred on the interfacial SNO, which substantiates that the linear band crossing (red dashed circle, $W_1$) forms a Weyl point. (g) The energy difference between the upper and lower bands shows the pair of Weyl points ($W_1$ and $W_2$) along the X-M path towards $\pm k_y$. (h) Equal and opposite Berry curvature peaks at $W_1$ and $W_2$ confirming the Weyl pair.}
    \label{fig:theory}
\end{figure*}
To obtain and analyze the octahedral rotations in the SNO layers, we performed a full relaxation of the heterostructure using VASP~\cite{Kresse1996} with a k-mesh of $7\times5\times1$ with a force convergence threshold of $10^{-4}$~eV/\AA. The structures before and after full relaxation are shown in Fig.~\ref{fig:theory}(a), \textcolor{black}{with the corresponding CIF files in the supporting information.}~\cite{supple} Upon analyzing the relaxed structure, we observe two major structural changes in the SNO layers. First, we find that the SNO octahedra undergo an out of phase rotation of $5.85^0$ along the c-axis, which corresponds to a glazer notation of $a^0a^0c^-$ (The octahedra are shown in Fig.~\ref{fig:theory}(b)). Secondly, there is a distortion of the octahedra at the interfacial SNO layer, as shown in Fig.~\ref{fig:theory}(c), where the central Nb atom is displaced by $0.46\ \textup{\AA}$ downward compared to the case of bulk SNO.
In addition, the interfacial structure exhibits a screw axis symmetry along $\vec{b}$~(see Fig.~\ref{fig:theory}(c)). Such a nonsymmorphic symmetry can allow the material to potentially host Dirac~\cite{Schoop2016,mohanta2021semi} or Weyl points.~\cite{PhysRevMaterials.4.124203,PhysRevMaterials.3.124204}\par
Next, we investigate the band structure of the relaxed system, which is shown in Fig.~\ref{fig:theory}(c). The SNO characters are projected onto the band structure and shown as blue dots. The band structure calculation was performed in the presence of spin orbit coupling with a hubbard $U =4.64\ eV$ for the Fe atoms. Upon close examination, we find that there are several linear band crossings present in the band structure near the Fermi level, shown by the dashed circles in Fig.~\ref{fig:theory}(c). 
However, to confirm their Dirac or Weyl nature, one must properly determine the topological properties. For this purpose, we extended our study to a bulk structure, keeping the octahedral rotation and distortion obtained from the fully relaxed heterostructure. We constructed a numerical tight binding model based on the hopping parameters extracted from DFT, using Wannier90~\cite{Pizzi_2020} and Quantum Espresso~\cite{Giannozzi_2009} code packages. The Berry curvatures for the individual bands were computed using the expression
\begin{equation} \label{obc}         
 \Omega^{\gamma}_{n,\alpha\beta} ({\vec  k}) = 2 \hbar   \sum_{n^\prime \neq n} \frac {{\rm Im}[ \langle u_{n{\vec  k}} | {v}_\alpha | u_{n^\prime{\vec  k}} \rangle  
\langle u_{n^\prime{\vec  k}} | v_\beta | u_{n{\vec  k}} \rangle]} 
{(\varepsilon_{n^\prime \vec k}-\varepsilon_{n  \vec k} )^2},
\end{equation}
where $v_{\alpha/\beta} =  \frac{1}{\hbar}{\partial {\cal H} }/ \partial k_{\alpha/\beta}$ is the velocity operator, $u_{m/n\ \vec{k}},$ and $\varepsilon_{m/n\ \vec{k}}$ are the eigenstates and eigenvalues for the $m^{th}$ and $n^{th}$ bands respectively. 
Interestingly, we found that all band crossings near the Fermi level, which may seem like Dirac points, as shown in Fig.~\ref{fig:theory}(e)(dashed circles) do not show any topological features. However, the point $W_1$ in the vicinity of X point (red dashed circle), is particularly interesting, which occurs at (0.5, 0.16, 0), and $\sim0.2\ eV$ above the Fermi level. A close examination of the 3D band structure as shown in Fig.~\ref{fig:theory}(f) reveals that the two bands near $W_1$ touch only at a single point and manifest large Berry curvature peaks, thus confirming its topological nature. Additionally, the opposite Berry curvatures of the upper and lower bands along with the two fold degeneracy confirm $W_1$ to be a Weyl point. In further analysis, we find that the topological partner $W_2$ of the point $W_1$ is located at (0.5,-0.16,0). The band touching points can be seen from the energy difference between the upper and lower bands near the X point, which is plotted in Fig.~\ref{fig:theory}(g). Together, they create a Weyl pair with exactly opposite Berry curvatures, as can be seen from Fig.~\ref{fig:theory}(h).  The various components of the Berry curvatures for the two bands near $W_1$ are shown in \textcolor{black}{Fig.~S7~\cite{supple}}. All three components show sharp peaks at $W_1$, while $\Omega^y_{xz}$ being an order of magnitude stronger than the rest. 
We note that there has been a theoretical study~\cite{PhysRevMaterials.7.075002} on bulk SNO, where it is shown that the $a^0a^0c^-$ rotation of the ideal octahedra can lead to the formation of Dirac points, unlike the distortion driven Weyl point seen in the present heterostructure.
It is important to highlight that until now, we have not considered any proximity magnetic field due to the LFO layers in the Hamiltonian, while computing the Berry curvatures. This signifies that the Weyl pairs occur purely due to the structural reorientation of the interfacial atoms and not as a result of the time-reversal symmetry  breaking induced by the proximity effect.
Furthermore, to account for the magnetic proximity effect induced by the LFO layers at the interface, we examined the energy separation between spin-polarized states obtained from the non-SOC band structure, as seen in \textcolor{black}{Fig.~S8~\cite{supple}}. The calculated values lie in the range of 0.02–0.06 eV. Since the time-reversal symmetry breaking can give rise to anomalous Hall effect, we calculated the anomalous Hall conductivity (AHC), considering an external magnetic field term in the Halmiltonian. The AHC was computed using the Kubo formula
$\sigma^{\gamma}_{\alpha \beta}   =  -\frac{e} { N_k V_c} \sum_{ \vec  k} \Omega^{\gamma}_{\alpha \beta} ({\vec  k}),$ where $V_c$ is the volume of the unit cell and $\Omega_{\alpha\beta}^\gamma$ is the Berry curvature. We found that the AHCs are in the range of -0.03 to 0.03 $\ \hbar/e\ \Omega^{-1}cm^{-1}$ in the presence of magnetic field splittings upto 0.06 eV. This is quite small compared to the typical values reported in other materials in the context of proximity induced anomalous Hall effect~\cite{PhysRevB.89.214406,Marfoua2022,Mesbahi2024}. 
In conclusion, our theoretical results indicate that the rotation and distortion of the SNO octahedra forms a Weyl pair near the E$_{f}$ protected by a screw axis symmetry, \textcolor{black}{which could contribute to the observed unconventional transport behaviour.}
\section{Summary}
Our study uncovers the emergence of Weyl states and related chiral transport in a structurally modulated SNO layer interfaced with antiferromagnetic insulating LFO, forming SNO/LFO bilayers. Temperature and field dependent transport measurements reveal a large, non-saturating linear MR, a nonlinear Hall response characteristic of multiband transport, field-induced reentrant metallic behavior, and a pronounced chiral-anomaly feature in the in-plane longitudinal MR ($B \parallel I$). These observations provide evidence for nontrivial electronic transport in SNO/LFO heterostructures. \textcolor{black}{Besides, a} derivative based analysis of Hall data further identifies a signature of an anomalous Hall contribution, likely originating from proximity effects induced by the interfacial LFO layers, which is otherwise masked by the multiband Hall response. First-principles density functional theory (DFT) calculations support the experimental findings by linking the nontrivial transport behavior to the emergence of a twofold-degenerate Weyl state induced by structural modulation in the SNO layer, where the NbO$_6$ octahedra of SNO in the bilayer undergo an out-of-phase $a^{0}a^{0}c^{-}$ type octahedral rotation accompanied by interfacial octahedral distortion. This is further confirmed by Berry curvature calculations, which show opposite sign Berry curvature peaks for the upper and lower bands, characteristic of a Weyl node. Our combined experimental and theoretical study establishes the realization of a strain and interfacial-distortion induced Weyl semimetallic phase in SNO/LFO bilayers, advancing the exploration of correlated topological states in $d$-electron-based complex oxide heterostructures.
\section{Experimental and Computational Methods}
\subsection {Sample Growth:}
Thin films of  SNO(9 nm)/LFO(5 nm), SNO(9 nm)/LFO(22 nm) bilayers \textcolor{blue}{as well as, bare SNO and LFO layers} were grown on (00$1$) oriented STO substrates using PLD, with a KrF excimer laser ($\lambda = 248~nm$), operating at a repetition rate of 4 Hz and a laser energy of 400~mJ. \textcolor{blue}{Sr$_{2}$Nb$_{2}$O$_{7}$ and LaFeO$_{3}$ targets were used for the growth of the respective layers}. The substrate temperature was maintained at 650$^{0}$~C for all depositions. SNO films were grown under a base pressure of 2$\times 10^{-6}$. In the case of the SNO/LFO bilayer, the LFO layer was initially deposited at 650$^{0}$~C, under an oxygen pressure of 2~mTorr, followed by deposition of the SNO layer at the same temperature under a base pressure of 2$\times 10^{-6}$.

\subsection{X-ray and Transport Characterization :}
The structural characterization of the films were examined using Rigaku SmartLab High-resolution X-ray diffraction (HR-XRD) with Cu Kα radiation source ($\lambda = 1.54059$~\AA).  Thin films were wire-bonded using $Al$ wires \textcolor{black}{for} electrical transport measurements. \textcolor{black}{Magnetotransport
measurements were performed in a \textcolor{blue}{CRYOGENIC} physical
property measurement system (PPMS). The magnetic field was applied both parallel and
perpendicular to the film surface.}\\
\subsection{DFT Methodology :}
The theoretical calculations in this work were performed using the following approach. In the case of structural relaxation calculation for SNO/LFO heterostructure, density functional theory calculations were performed using the Vienna Ab initio Simulation Package (VASP)~\cite{kresse1993ab,kresse1996efficient},  employing the generalized gradient approximation in the form of the Perdew-Burke-Ernzerhof (PBE) functional~\cite{perdew1996generalized}. The structures were relaxed until the residual forces on each atom were less than $10^{-4}$~eV/\AA. On the other hand, the topological properties were evaluated using tight-binding models constructed from hopping parameters obtained from maximally localized Wannier functions generated with Wannier90~\cite{mostofi2008wannier90}. The Wannier functions were converged with a localization tolerance of $10^{-10}$~\AA. The underlying self-consistent field calculations were performed using Quantum ESPRESSO~\cite{giannozzi2009quantum} with a plane-wave energy cutoff of 200~Ry and an electronic energy convergence threshold of $10^{-9}$~eV. 
\section*{Acknowledgment}
D.S. acknowledges the funding from Max Planck Partner Group and SERB, Government of India (Grant No. CRG/2019/005144). S.I. and D.S thanks G.Tripathy, S.D Mohanty,and N.Mohanta, for many valuable insights and fruitful discussions. S.I. thanks Tapas, Biswajit  for assistance during the measurements and K. Hayagreev for helpful support during the preparation of the figures.
\section*{Conflicts  of Interest}
The authors declare no conflicts of interest.
\section*{Data Availability Statement}
The data that support the findings of this study are available from the corresponding author upon reasonable request
\bibliography{Bibilography}

@article{lv2021experimental,
  title={Experimental perspective on three-dimensional topological semimetals},
  author={Lv, BQ and Qian, T and Ding, H},
  journal={Reviews of Modern Physics},
  volume={93},
  number={2},
  pages={025002},
  year={2021},
  publisher={APS},
  url= {https://link.aps.org/doi/10.1103/RevModPhys.93.025002}
}

@article{coutinho2017structural,
  title={Structural, vibrational and magnetic properties of the orthoferrites {LaFeO$_3$} and {YFeO$_3$}: A comparative study},
  author={Coutinho, PV and Cunha, F and Barrozo, Petrucio},
  journal={Solid State Communications},
  volume={252},
  pages={59-63},
  year={2017},
  publisher={Elsevier},
  url = {https://www.sciencedirect.com/science/article/pii/S0038109817300273?via%3Dihub}  
}

@article{panda2025tuning,
  title={Tuning of magnetic anisotropy and metal-insulator transition via heterointerface coupling in {LaFeO$_3$/SrRuO$_3$} bilayers},
  author={Panda, Soumyakanta and Das, Malaya Kumar and Mohapatra, Rasmi Ranjan and Nayak, Debashree and Padhan, Prahallad and Samal, D and Mohapatra, N},
  journal={Physical Review B},
  volume={112},
  number={5},
  pages={054428},
  year={2025},
  publisher={APS},
  url = {https://doi.org/10.1103/svc8-dy9j}
}

@article{asmara2017tunable,
  title={Tunable and low-loss correlated plasmons in Mott-like insulating oxides},
  author={Asmara, Teguh Citra and Wan, Dongyang and Zhao, Yongliang and Majidi, Muhammad Aziz and Nelson, Christopher T and Scott, Mary C and Cai, Yao and Yan, Bixing and Schmidt, Daniel and Yang, Ming and others},
  journal={Nature Communications},
  volume={8},
  number={1},
  pages={15271},
  year={2017},
  url = {https://www.nature.com/articles/ncomms15271.pdf},
  publisher={Nature Publishing Group UK London}
}

@article{chen2013significant,
  title={Significant ferrimagnetisms observed in superlattice composed of antiferromagnetic {LaFeO$_3$} and {YMnO$_3$}},
  author={Chen, YB and Zhou, Jian and Zhang, Shan-Tao and Wu, Fei-Xiang and Yao, Shu-Hua and Gu, Zheng-Bin and Wu, Di and Chen, Yan-Feng},
  journal={Applied Physics Letters},
  volume={102},
  number={4},
  year={2013},
  publisher={AIP Publishing},
  url = {https://doi.org/10.1063/1.4789753}
}

@article{narang2021topology,
  title={The topology of electronic band structures},
  author={Narang, Prineha and Garcia, Christina AC and Felser, Claudia},
  journal={Nature Materials},
  volume={20},
  number={3},
  pages={293-300},
  year={2021},
  publisher={Nature Publishing Group UK London},
  url  = {https://www.nature.com/articles/s41563-020-00820-4}
}

@article{kumar2020topological,
  title={Topological quantum materials from the viewpoint of chemistry},
  author={Kumar, Nitesh and Guin, Satya N and Manna, Kaustuv and Shekhar, Chandra and Felser, Claudia},
  journal={Chemical Reviews},
  volume={121},
  number={5},
  pages={2780-2815},
  year={2020},
  publisher={ACS Publications},
  url = {https://pubs.acs.org/doi/10.1021/acs.chemrev.0c00732}
}

@article{hu2019transport,
  title={Transport of topological semimetals},
  author={Hu, Jin and Xu, Su-Yang and Ni, Ni and Mao, Zhiqiang},
  journal={Annual Review of Materials Research},
  volume={49},
  number={1},
  pages={207-252},
  year={2019},
  publisher={Annual Reviews},
  url = {https://www.annualreviews.org/content/journals/10.1146/annurev-matsci-070218-010023}
}

@article{liu2014discovery,
  title={Discovery of a three-dimensional topological {Dirac} semimetal, {Na$_3$Bi}},
  author={Liu, ZK and Zhou, Bo and Zhang, Yong and Wang, ZJ and Weng, HM and Prabhakaran, Dharmalingam and Mo, S-K and Shen, ZX and Fang, Zhong and Dai, Xi and others},
  journal={Science},
  volume={343},
  number={6173},
  pages={864-867},
  year={2014},
  publisher={American Association for the Advancement of Science},
  url = {https://www.science.org/doi/10.1126/science.1245085}
}

@article{lv2015experimental,
  title={Experimental discovery of {Weyl} semimetal {TaAs}},
  author={Lv, BQ and Weng, HM and Fu, BB and Wang, X Ps and Miao, Hu and Ma, Junzhang and Richard, P and Huang, XC and Zhao, LX and Chen, GF and others},
  journal={Physical Review X},
  volume={5},
  number={3},
  pages={031013},
  year={2015},
  publisher={APS},
  url = {https://journals.aps.org/prx/abstract/10.1103/PhysRevX.5.031013}
}

@article{yang2015weyl,
  title={{Weyl} semimetal phase in the non-centrosymmetric compound {TaAs}},
  author={Yang, LX and Liu, ZK and Sun, Yan and Peng, Han and Yang, HF and Zhang, Teng and Zhou, Beatrice and Zhang, Yi and Guo, YF and Rahn, Marein and others},
  journal={Nature Physics},
  volume={11},
  number={9},
  pages={728-732},
  year={2015},
  publisher={Nature Publishing Group UK London},
  url = {https://www.nature.com/articles/nphys3425}
}

@article{meena2025nonsymmorphic,
  title={Nonsymmorphic symmetry protected hourglass {Dirac} chain topology and conventional superconductivity in {ZrIrGe}},
  author={Meena, Pavan Kumar and Samanta, Dibyendu and Srivastava, Shashank and Manna, Poulami and Ghosh, Sudeep Kumar and Singh, Ravi Prakash},
  journal={Physical Review B},
  volume={112},
  number={14},
  pages={144515},
  year={2025},
  publisher={APS},
  url = {https://journals.aps.org/prb/abstract/10.1103/dch7-3f7y}
}

@article{yang2014emergent,
  title={Emergent topological phenomena in thin films of {Pyrochlore Iridates}},
  author={Yang, Bohm-Jung and Nagaosa, Naoto},
  journal={Physical Review Letters},
  volume={112},
  number={24},
  pages={246402},
  year={2014},
  publisher={APS},
  url = {https://journals.aps.org/prl/abstract/10.1103/PhysRevLett.112.246402}
}

@article{rosendal2023octahedral,
  title = {Octahedral distortions in {SrNbO$_3$}: unraveling the structure-property relation},
  author={Rosendal, Victor and Brito, Walber H and Radovic, Milan and Chikina, Alla and Brandbyge, Mads and Pryds, Nini and Petersen, Dirch H},
  journal={Physical Review Materials},
  volume={7},
  number={7},
  pages={075002},
  year={2023},
  url = {https://doi.org/10.1103/PhysRevMaterials.7.075002},
  publisher={APS}
}

@article{chikina2025intrinsic,
  title={{Intrinsic} three-dimensional topology in  {SrNbO$_3$} films},
  author={Chikina, Alla and Rosendal, Victor and Li, Hang and Skoropata, Elizabeth and Guedes, Eduardo B and Caputo, Marco and Plumb, NC and Shi, M and Petersen, DH and Brandbyge, M and others},
  journal={Physical Review B},
  volume={111},
  number={15},
  pages={155146},
  year={2025},
  publisher={APS},
  url = {https://journals.aps.org/prb/abstract/10.1103/PhysRevB.111.155146}
}

@article{fujioka2019strong,
  title={Strong-correlation induced high-mobility electrons in {Dirac} semimetal of perovskite oxide},
  author={Fujioka, J and Yamada, R and Kawamura, M and Sakai, S and Hirayama, M and Arita, R and Okawa, T and Hashizume, D and Hoshino, M and Tokura, Y},
  journal={Nature Communications},
  volume={10},
  number={1},
  pages={362},
  year={2019},
  publisher={Nature Publishing Group UK London},
  url = {https://www.nature.com/articles/s41467-018-08149-y}
}

@article{takahashi2018anomalous,
  title={Anomalous {Hall} effect derived from multiple {Weyl} nodes in high-mobility {EuTiO$_3$} films},
  author={Takahashi, Kei S and Ishizuka, Hiroaki and Murata, Tomoki and Wang, Qing Y and Tokura, Yoshinori and Nagaosa, Naoto and Kawasaki, Masashi},
  journal={Science Advances},
  volume={4},
  number={7},
  pages={eaar7880},
  year={2018},
  publisher={American Association for the Advancement of Science},
  url = {https://www.science.org/doi/10.1126/sciadv.aar7880}
}

@article{zhang2019catalogue,
  title={Catalogue of topological electronic materials},
  author={Zhang, Tiantian and Jiang, Yi and Song, Zhida and Huang, He and He, Yuqing and Fang, Zhong and Weng, Hongming and Fang, Chen},
  journal={Nature},
  volume={566},
  number={7745},
  pages={475-479},
  year={2019},
  publisher={Nature Publishing Group UK London},
  url = {https://www.nature.com/articles/s41586-019-0944-6}
}

@article{singha2017large,
  title={Large nonsaturating magnetoresistance and signature of nondegenerate {Dirac} nodes in {ZrSiS}},
  author={Singha, Ratnadwip and Pariari, Arnab Kumar and Satpati, Biswarup and Mandal, Prabhat},
  journal={Proceedings of the National Academy of Sciences},
  volume={114},
  number={10},
  pages={2468-2473},
  year={2017},
  publisher={National Academy of Sciences},
  url = {https://www.pnas.org/doi/10.1073/pnas.1618004114}
}

@article{leahy2018nonsaturating,
  title={Nonsaturating large magnetoresistance in semimetals},
  author={Leahy, Ian A and Lin, Yu-Ping and Siegfried, Peter E and Treglia, Andrew C and Song, Justin CW and Nandkishore, Rahul M and Lee, Minhyea},
  journal={Proceedings of the National Academy of Sciences},
  volume={115},
  number={42},
  pages={10570-10575},
  year={2018},
  publisher={National Academy of Sciences},
  url = {https://www.pnas.org/doi/10.1073/pnas.1808747115}
}

@article{xu2015observation,
  title={Observation of {Fermi} arc surface states in a topological metal},
  author={Xu, Su-Yang and Liu, Chang and Kushwaha, Satya K and Sankar, Raman and Krizan, Jason W and Belopolski, Ilya and Neupane, Madhab and Bian, Guang and Alidoust, Nasser and Chang, Tay-Rong and others},
  journal={Science},
  volume={347},
  number={6219},
  pages={294-298},
  year={2015},
  publisher={American Association for the Advancement of Science},
  url = {https://www.science.org/doi/10.1126/science.1256742}
}

@article{wan2011topological,
  title={Topological semimetal and {Fermi}-arc surface states in the electronic structure of pyrochlore iridates},
  author={Wan, Xiangang and Turner, Ari M and Vishwanath, Ashvin and Savrasov, Sergey Y},
  journal={Physical Review B},
  volume={83},
  number={20},
  pages={205101},
  year={2011},
  publisher={APS},
  url = {https://journals.aps.org/prb/abstract/10.1103/PhysRevB.83.205101}
}

@article{thakur2020intrinsic,
  title={Intrinsic anomalous Hall effect in {Ni}-substituted magnetic {Weyl} semimetal {Co$_3$Sn$_2$S$_2$}},
  author={Thakur, Gohil S and Vir, Praveen and Guin, Satya N and Shekhar, Chandra and Weihrich, Richard and Sun, Yan and Kumar, Nitesh and Felser, Claudia},
  journal={Chemistry of Materials},
  volume={32},
  number={4},
  pages={1612-1617},
  year={2020},
  publisher={ACS Publications},
  url = {https://pubs.acs.org/doi/10.1021/acs.chemmater.9b05009}
}

@article{zyuzin2017magnetotransport,
  title={Magnetotransport of {Weyl} semimetals due to the chiral anomaly},
  author={Zyuzin, Vladimir A},
  journal={Physical Review B},
  volume={95},
  number={24},
  pages={245128},
  year={2017},
  publisher={APS},
  url = {https://journals.aps.org/prb/abstract/10.1103/PhysRevB.95.245128}
}

@article{zyuzin2012topological,
  title={Topological response in {Weyl} semimetals and the chiral anomaly},
  author={Zyuzin, AA and Burkov, AA},
  journal={Physical Review B},
  volume={86},
  number={11},
  pages={115133},
  year={2012},
  publisher={APS},
  url = {https://journals.aps.org/prb/abstract/10.1103/PhysRevB.86.115133}
}

@article{chen2025extremely,
  title={Extremely large magnetoresistance and {Shubnikov-de Haas} oscillations in topological semimetal {FeGe$_2$}},
  author={Chen, Yuan-Qiao and Wu, Wei-Bin and Du, Mei and Fan, Yi-Wei and He, Ya-Xun and Zhang, Jia-Ying and Xiao, Qi-Ling and Li, Zhe and Ge, Jun-Yi},
  journal={Physical Review B},
  volume={112},
  number={11},
  pages={115149},
  year={2025},
  url = { https://doi.org/10.1103/6ycd-k56z},
  publisher={APS}
}

@article{masuko2019strain,
  title={Strain-engineering of charge transport in the correlated {Dirac} semimetal of perovskite {CaIrO$_3$} thin films},
  author={Masuko, M and Fujioka, J and Nakamura, M and Kawasaki, M and Tokura, Y},
  journal={APL Materials},
  volume={7},
  number={8},
  year={2019},
  publisher={AIP Publishing},
  url = {https://pubs.aip.org/aip/apm/article/7/8/081115/123154/Strain-engineering-of-charge-transport-in-the}
}

@article{chen2016topological,
  title={Topological crystalline semimetals in nonsymmorphic lattices},
  author={Chen, Yige and Kim, Heung-Sik and Kee, Hae-Young},
  journal={Physical Review B},
  volume={93},
  number={15},
  pages={155140},
  year={2016},
  publisher={APS},
  url = {https://journals.aps.org/prb/abstract/10.1103/PhysRevB.93.155140}
}

@article{sun2015topological,
  title={Topological surface states and {Fermi} arcs of the noncentrosymmetric {Weyl} semimetals {TaAs}, {TaP}, {NbAs}, and {NbP}},
  author={Sun, Yan and Wu, Shu-Chun and Yan, Binghai},
  journal={Physical Review B},
  volume={92},
  number={11},
  pages={115428},
  year={2015},
  publisher={APS},
  url = {https://journals.aps.org/prb/abstract/10.1103/PhysRevB.92.115428}
}

@article{souma2016direct,
  title={Direct observation of nonequivalent {Fermi}-arc states of opposite surfaces in the noncentrosymmetric {Weyl} semimetal {NbP}},
  author={Souma, S and Wang, Zhiwei and Kotaka, H and Sato, T and Nakayama, K and Tanaka, Y and Kimizuka, H and Takahashi, T and Yamauchi, K and Oguchi, T and others},
  journal={Physical Review B},
  volume={93},
  number={16},
  pages={161112},
  year={2016},
  publisher={APS},
  url = {https://journals.aps.org/prb/abstract/10.1103/PhysRevB.93.161112}
}

@article{li2017evidence,
  title={Evidence for topological type-{II} {Weyl} semimetal {WTe$_2$}},
  author={Li, Peng and Wen, Yan and He, Xin and Zhang, Qiang and Xia, Chuan and Yu, Zhi-Ming and Yang, Shengyuan A and Zhu, Zhiyong and Alshareef, Husam N and Zhang, Xi-Xiang},
  journal={Nature Communications},
  volume={8},
  number={1},
  pages={2150},
  year={2017},
  publisher={Nature Publishing Group UK London},
  url = {https://www.nature.com/articles/s41467-017-02237-1}
}

@article{deng2016experimental,
  title={Experimental observation of topological Fermi arcs in type-{II} {Weyl} semimetal {MoTe$_2$}},
  author={Deng, Ke and Wan, Guoliang and Deng, Peng and Zhang, Kenan and Ding, Shijie and Wang, Eryin and Yan, Mingzhe and Huang, Huaqing and Zhang, Hongyun and Xu, Zhilin and others},
  journal={Nature Physics},
  volume={12},
  number={12},
  pages={1105-1110},
  year={2016},
  publisher={Nature Publishing Group UK London},
  url = {https://www.nature.com/articles/nphys3871}
}

@article{xu2015experimental,
  title={Experimental discovery of a topological {Weyl} semimetal state in {TaP}},
  author={Xu, Su-Yang and Belopolski, Ilya and Sanchez, Daniel S and Zhang, Chenglong and Chang, Guoqing and Guo, Cheng and Bian, Guang and Yuan, Zhujun and Lu, Hong and Chang, Tay-Rong and others},
  journal={Science Advances},
  volume={1},
  number={10},
  pages={e1501092},
  year={2015},
  publisher={American Association for the Advancement of Science},
  url = {https://www.science.org/doi/10.1126/sciadv.1501092}
}

@article{oka2015intrinsic,
  title={Intrinsic high electrical conductivity of stoichiometric {SrNbO$_3$} epitaxial thin films},
  author={Oka, Daichi and Hirose, Yasushi and Nakao, Shoichiro and Fukumura, Tomoteru and Hasegawa, Tetsuya},
  journal={Physical Review B},
  year={2015},
  url= {https://doi.org/10.1103/PhysRevB.92.205102}
}

@article{kresse1993ab,
  title={Ab initio molecular dynamics for liquid metals},
  author={Kresse, Georg and Hafner, J{\"u}rgen},
  journal={Physical Review B},
  volume={47},
  number={1},
  pages={558},
  year={1993},
  publisher={APS},
  url = { https://doi.org/10.1103/PhysRevB.47.558}
}

@article{kresse1996efficient,
  title={Efficient iterative schemes for ab initio total-energy calculations using a plane-wave basis set},
  author={Kresse, Georg and Furthm{\"u}ller, J{\"u}rgen},
  journal={Physical Review B},
  volume={54},
  number={16},
  pages={11169},
  year={1996},
  publisher={APS},
  url = {https://doi.org/10.1103/PhysRevB.54.11169}
}

@article{perdew1996generalized,
  title={Generalized gradient approximation made simple},
  author={Perdew, John P and Burke, Kieron and Ernzerhof, Matthias},
  journal={Physical Review Letters},
  volume={77},
  number={18},
  pages={3865},
  year={1996},
  url = {https://doi.org/10.1103/PhysRevLett.77.3865},
  publisher={APS}
}

@article{mostofi2008wannier90,
  title={wannier90: A tool for obtaining maximally-localised Wannier functions},
  author={Mostofi, Arash A and Yates, Jonathan R and Lee, Young-Su and Souza, Ivo and Vanderbilt, David and Marzari, Nicola},
  journal={Computer physics communications},
  volume={178},
  number={9},
  pages={685-699},
  year={2008},
  url = {https://doi.org/10.1016/j.cpc.2007.11.016},
  publisher={Elsevier}
}

@article{giannozzi2009quantum,
  title={QUANTUM ESPRESSO: a modular and open-source software project for quantum simulations of materials},
  author={Giannozzi, Paolo and Baroni, Stefano and Bonini, Nicola and Calandra, Matteo and Car, Roberto and Cavazzoni, Carlo and Ceresoli, Davide and Chiarotti, Guido L and Cococcioni, Matteo and Dabo, Ismaila and others},
  journal={Journal of physics: Condensed matter},
  volume={21},
  number={39},
  pages={395502},
  year={2009},
  url = {https://iopscience.iop.org/article/10.1088/0953-8984/21/39/395502}
}

@article{yan2017topological,
  title={Topological materials: {Weyl} semimetals},
  author={Yan, Binghai and Felser, Claudia},
  journal={Annual Review of Condensed Matter Physics},
  volume={8},
  pages={337-354},
  year={2017},
  publisher={Annual Reviews},
  url = {https://www.annualreviews.org/content/journals/10.1146/annurev-conmatphys-031016-025458}
}

@article{young2012dirac,
  title={Dirac semimetal in three dimensions},
  author={Young, Steve M and Zaheer, Saad and Teo, Jeffrey CY and Kane, Charles L and Mele, Eugene J and Rappe, Andrew M},
  journal={Physical Review Letters},
  volume={108},
  number={14},
  pages={140405},
  year={2012},
  publisher={APS},
  url = {https://journals.aps.org/prl/abstract/10.1103/PhysRevLett.108.140405}
}

@article{wang2017quantum,
  title={Quantum transport in {Dirac} and {Weyl} semimetals: a review},
  author={Wang, Shuo and Lin, Ben-Chuan and Wang, An-Qi and Yu, Da-Peng and Liao, Zhi-Min},
  journal={Advances in Physics: X},
  volume={2},
  number={3},
  pages={518-544},
  year={2017},
  publisher={Taylor \& Francis},
  url = {https://doi.org/10.1080/23746149.2017.1327329}
}

@article{samal2016molecular,
  title={Molecular beam epitaxy of three-dimensional {Dirac} material {Sr$_3$PbO}},
  author={Samal, D and Nakamura, H and Takagi, H},
  journal={APL Materials},
  volume={4},
  number={7},
  year={2016},
  url = {https://pubs.aip.org/aip/apm/article/4/7/076101/120966/Molecular-beam-epitaxy-of-three-dimensional-Dirac},
  publisher={AIP Publishing}
}

@article{nakamura2020robust,
  title={Robust weak antilocalization due to spin-orbital entanglement in {Dirac} material {Sr$_3$SnO}},
  author={Nakamura, H and Huang, D and Merz, J and Khalaf, E and Ostrovsky, P and Yaresko, A and Samal, D and Takagi, H},
  journal={Nature Communications},
  volume={11},
  number={1},
  pages={1161},
  year={2020},
  publisher={Nature Publishing Group UK London},
  url = {https://www.nature.com/articles/s41467-020-14900-1}
}

@article{konig2007quantum,
  title={Quantum spin Hall insulator state in {HgTe} quantum wells},
  author={Konig, Markus and Wiedmann, Steffen and Brune, Christoph and Roth, Andreas and Buhmann, Hartmut and Molenkamp, Laurens W and Qi, Xiao-Liang and Zhang, Shou-Cheng},
  journal={Science},
  volume={318},
  number={5851},
  pages={766-770},
  year={2007},
  publisher={American Association for the Advancement of Science},
  url = {https://www.science.org/doi/10.1126/science.1148047}
}

@article{bernevig2006quantum,
  title={Quantum spin {Hall} effect and topological phase transition in {HgTe} quantum wells},
  author={Bernevig, B Andrei and Hughes, Taylor L and Zhang, Shou-Cheng},
  journal={Science},
  volume={314},
  number={5806},
  pages={1757-1761},
  year={2006},
  publisher={American Association for the Advancement of Science},
  url = {https://www.science.org/doi/10.1126/science.1133734}
}

@article{zhang2009topological,
  title={{Topological insulators} in {Bi$_2$Se$_3$}, {Bi$_2$Te$_3$} and {Sb$_2$Te$_3$} with a single {Dirac} cone on the surface},
  author={Zhang, Haijun and Liu, Chao-Xing and Qi, Xiao-Liang and Dai, Xi and Fang, Zhong and Zhang, Shou-Cheng},
  journal={Nature Physics},
  volume={5},
  number={6},
  pages={438-442},
  year={2009},
  url = {https://www.nature.com/articles/nphys1270},
  publisher={Nature Publishing Group UK London}
}

@article{raghu2008topological,
  title={Topological mott insulators},
  author={Raghu, Srinivas and Qi, Xiao-Liang and Honerkamp, Carsten and Zhang, Shou-Cheng},
  journal={Physical Review Letters},
  volume={100},
  number={15},
  pages={156401},
  year={2008},
  url = {https://doi.org/10.1103/PhysRevLett.100.156401},
  publisher={APS}
}

@article{liu2020robust,
  title={Robust axion insulator and Chern insulator phases in a two-dimensional antiferromagnetic topological insulator},
  author={Liu, Chang and Wang, Yongchao and Li, Hao and Wu, Yang and Li, Yaoxin and Li, Jiaheng and He, Ke and Xu, Yong and Zhang, Jinsong and Wang, Yayu},
  journal={Nature Materials},
  volume={19},
  number={5},
  pages={522-527},
  year={2020},
  url = {https://www.nature.com/articles/s41563-019-0573-3},
  publisher={Nature Publishing Group UK London}
}

@article{mandal2023topological,
  title={Topological superconductors from a materials perspective},
  author={Mandal, Manasi and Drucker, Nathan C and Siriviboon, Phum and Nguyen, Thanh and Boonkird, Artittaya and Lamichhane, Tej Nath and Okabe, Ryotaro and Chotrattanapituk, Abhijatmedhi and Li, Mingda},
  journal={Chemistry of Materials},
  volume={35},
  number={16},
  pages={6184-6200},
  year={2023},
  url = {https://pubs.acs.org/doi/10.1021/acs.chemmater.3c00713},
  publisher={ACS Publications}
}

@article{huang2021review,
 title = {A review of some new perspectives on the theory of superconducting {Sr$_2$RuO$_4$}},
  author={Huang, Wen},
  journal={Chinese Physics B},
  volume={30},
  number={10},
  pages={107403},
  year={2021},
  publisher={Chinese Physical Society and IOP Publishing Ltd},
  url = {https://iopscience.iop.org/article/10.1088/1674-1056/ac2488}
}

@article{fang2019discovery,
  title = {Discovery of superconductivity in 2M {WS$_2$} with possible topological surface states},
  author={Fang, Yuqiang and Pan, Jie and Zhang, Dongqin and Wang, Dong and Hirose, Hishiro T and Terashima, Taichi and Uji, Shinya and Yuan, Yonghao and Li, Wei and Tian, Zhen and others},
  journal={Advanced Materials},
  volume={31},
  number={30},
  pages={1901942},
  year={2019},
  publisher={Wiley Online Library},
  url = {https://advanced.onlinelibrary.wiley.com/doi/10.1002/adma.201901942?utm_medium=article&utm_source=researchgate.net},
}

@article{yang2011quantum,
  title = {{Quantum Hall} effects in a {Weyl} semimetal: Possible application in pyrochlore iridates},
  author={Yang, Kai-Yu and Lu, Yuan-Ming and Ran, Ying},
  journal={Physical Review B},
  volume={84},
  number={7},
  pages={075129},
  year={2011},
  publisher={APS},
  url = {https://journals.aps.org/prb/abstract/10.1103/PhysRevB.84.075129}
}

@article{schoop2016dirac,
  title = {{Dirac} cone protected by non-symmorphic symmetry and three-dimensional {Dirac} line node in {ZrSiS}},
  author={Schoop, Leslie M and Ali, Mazhar N and Stra{\ss}er, Carola and Topp, Andreas and Varykhalov, Andrei and Marchenko, Dmitry and Duppel, Viola and Parkin, Stuart SP and Lotsch, Bettina V and Ast, Christian R},
  journal={Nature Communications},
  volume={7},
  number={1},
  pages={11696},
  year={2016},
  url = {https://www.nature.com/articles/ncomms11696},
  publisher={Nature Publishing Group UK London}
}

@article{li2018nonsymmorphic,
  title = {Nonsymmorphic-symmetry-protected hourglass {Dirac} loop, nodal line, and {Dirac} point in bulk and monolayer {X$_3$SiTe$_6$} (X = {Ta}, {Nb})},
  author={Li, Si and Liu, Ying and Wang, Shan-Shan and Yu, Zhi-Ming and Guan, Shan and Sheng, Xian-Lei and Yao, Yugui and Yang, Shengyuan A},
  journal={Physical Review B},
  volume={97},
  number={4},
  pages={045131},
  year={2018},
  publisher={APS},
  url = {https://journals.aps.org/prb/abstract/10.1103/PhysRevB.97.045131}
}

@article{ghosh2023spontaneous,
 title = {Spontaneous {Hall} effect in the magnetic {Weyl} semimetallic {Eu$_2$Ir$_2$O$_7$} (111) thin films},
  author={Ghosh, Mithun and Samal, D and Kumar, PS},
  journal={Applied Physics Letters},
  volume={123},
  number={21},
  year={2023},
  publisher={AIP Publishing},
  url = {https://doi.org/10.1063/5.0172127}
}

@article{li2016chiral,
  title={Chiral magnetic effect in {ZrTe}$_5$},
  author={Li, Qiang and Kharzeev, Dmitri E and Zhang, Cheng and Huang, Yuan and Pletikosi{\'c}, I and Fedorov, AV and Zhong, RD and Schneeloch, JA and Gu, GD and Valla, T},
  journal={Nature Physics},
  volume={12},
  number={6},
  pages={550-554},
  year={2016},
  url = {https://www.nature.com/articles/nphys3648},
  publisher={Nature Publishing Group UK London}
}

@article{li2016negative,
  title={Negative magnetoresistance in {Cd}$_3${As}$_2$},
  author={Li, Hui and He, Hongtao and Lu, Hai-Zhou and Zhang, Huachen and Liu, Hongchao and Ma, Rong and Fan, Zhiyong and Shen, Shun-Qing and Wang, Jiannong},
  journal={Nature Communications},
  volume={7},
  number={1},
  pages={10301},
  year={2016},
  url = {https://www.nature.com/articles/ncomms10301},
  publisher={Nature Publishing Group UK London}
}

@article{ong2021experimental,
  title = {Experimental signatures of the chiral anomaly in {Dirac}-{Weyl} semimetals},
  author={Ong, NP and Liang, Sihang},
  journal={Nature Reviews Physics},
  volume={3},
  number={6},
  pages={394-404},
  year={2021},
  url = {https://www.nature.com/articles/s42254-021-00310-9},
  publisher={Nature Publishing Group UK London}
}

@article{ok2021correlated,
 title = {Correlated oxide {Dirac} semimetal in the extreme quantum limit},
  author={Ok, Jong Mok and Mohanta, Narayan and Zhang, Jie and Yoon, Sangmoon and Okamoto, Satoshi and Choi, Eun Sang and Zhou, Hua and Briggeman, Megan and Irvin, Patrick and Lupini, Andrew R and others},
  journal={Science Advances},
  volume={7},
  number={38},
  pages={eabf9631},
  year={2021},
  url = {https://www.science.org/doi/10.1126/sciadv.abf9631},
  publisher={American Association for the Advancement of Science}
}

@article{neupane2014observation,
  title = {Observation of a three-dimensional topological {Dirac} semimetal phase in high-mobility {Cd$_3$As$_2$}},
  author={Neupane, Madhab and Xu, Su-Yang and Sankar, Raman and Alidoust, Nasser and Bian, Guang and Liu, Chang and Belopolski, Ilya and Chang, Tay-Rong and Jeng, Horng-Tay and Lin, Hsin and others},
  journal={Nature Communications},
  volume={5},
  number={1},
  pages={3786},
  year={2014},
  url = {https://www.nature.com/articles/ncomms4786},
  publisher={Nature Publishing Group UK London}
}

@article{mohanta2021semi,
  title = {Semi-{Dirac} and {Weyl} fermions in transition metal oxides},
  author={Mohanta, Narayan and Ok, Jong Mok and Zhang, Jie and Miao, Hu and Dagotto, Elbio and Lee, Ho Nyung and Okamoto, Satoshi},
  journal={Physical Review B},
  volume={104},
  number={23},
  pages={235121},
  year={2021},
  url = { https://doi.org/10.1103/PhysRevB.104.235121},
  publisher={APS}
}

@article{zhang2021extremely,
  title={Extremely large magnetoresistance in high-mobility {SrNbO}$_3$/{SrTiO}$_3$ heterostructures},
  author={Zhang, Jie and Ok, Jong Mok and Pai, Yun-Yi and Lapano, Jason and Skoropata, Elizabeth and Mazza, Alessandro R and Li, Haoxiang and Huon, Amanda and Yoon, Sangmoon and Lawrie, Benjamin and others},
  journal={Physical Review B},
  volume={104},
  number={16},
  pages={L161404},
  year={2021},
  url = {https://doi.org/10.1103/PhysRevB.104.L161404},
  publisher={APS}
}

@article{okuma2024large,
  title={Large Rashba parameter for 4d strongly correlated perovskite oxide {SrNbO}$_3$ ultrathin films},
  author={Okuma, Hikaru and Katayama, Yumiko and Ueno, Kazunori},
  journal={Physical Review Materials},
  volume={8},
  number={1},
  pages={015001},
  year={2024},
  publisher={APS},
  url =  {https://doi.org/10.1103/PhysRevMaterials.8.015001}
}

@article{liu2016strain,
  title={Strain-induced nonsymmorphic symmetry breaking and removal of {Dirac} semimetallic nodal line in an orthoperovskite iridate},
  author={Liu, Jian and Kriegner, Dominik and Horak, L and Puggioni, D and Rayan Serrao, C and Chen, R and Yi, D and Frontera, C and Holy, V and Vishwanath, A and others},
  journal={Physical Review B},
  volume={93},
  number={8},
  pages={085118},
  year={2016},
  url = {https://doi.org/10.1103/PhysRevB.93.085118},
  publisher={APS}
}

@article{huang2015observation,
  title = {{Observation} of the chiral-anomaly-induced negative magnetoresistance in 3D {Weyl} semimetal {TaAs}},
  author={Huang, Xiaochun and Zhao, Lingxiao and Long, Yujia and Wang, Peipei and Chen, Dong and Yang, Zhanhai and Liang, Hui and Xue, Mianqi and Weng, Hongming and Fang, Zhong and others},
  journal={Physical Review X},
  volume={5},
  number={3},
  pages={031023},
  year={2015},
  url = {https://doi.org/10.1103/PhysRevX.5.031023},
  publisher={APS}
}

@article{armitage2018weyl,
  title = {{Weyl} and {Dirac} semimetals in three-dimensional solids},
  author={Armitage, N Peter and Mele, Eugene J and Vishwanath, Ashvin},
  journal={Reviews of Modern Physics},
  volume={90},
  number={1},
  pages={015001},
  year={2018},
  url = {https://doi.org/10.1103/RevModPhys.90.015001},
  publisher={APS}
}

@article{li2015giant,
  title = {{Giant} negative magnetoresistance induced by the chiral anomaly in individual {Cd$_3$As$_2$} nanowires},
  author={Li, Cai-Zhen and Wang, Li-Xian and Liu, Haiwen and Wang, Jian and Liao, Zhi-Min and Yu, Da-Peng},
  journal={Nature Communications},
  volume={6},
  number={1},
  pages={10137},
  year={2015},
  url = {https://www.nature.com/articles/ncomms10137},
  publisher={Nature Publishing Group UK London}
}

@article{tafti2016resistivity,
  title = {{Resistivity} plateau and extreme magnetoresistance in {LaSb}},
  author={Tafti, FF and Gibson, QD and Kushwaha, SK and Haldolaarachchige, Neel and Cava, RJ},
  journal={Nature Physics},
  volume={12},
  number={3},
  pages={272-277},
  year={2016},
  url = {https://www.nature.com/articles/nphys3581},
  publisher={Nature Publishing Group UK London}
}

@article{xiong2015evidence,
 title = {Evidence for the chiral anomaly in the {Dirac} semimetal {Na$_3$Bi}},
  author={Xiong, Jun and Kushwaha, Satya K and Liang, Tian and Krizan, Jason W and Hirschberger, Max and Wang, Wudi and Cava, Robert Joseph and Ong, Nai Phuan},
  journal={Science},
  volume={350},
  number={6259},
  pages={413-416},
  year={2015},
  url = {https://www.science.org/doi/10.1126/science.aac6089},
  publisher={American Association for the Advancement of Science}
}

@article{zhang2025inadequacy,
  title={The inadequacy of the $\rho$-T curve for phase transitions in the presence of magnetic fields},
  author={Zhang, Shengnan and Fang, Zhong and Weng, Hongming and Wu, Quansheng},
  journal={The Innovation},
  volume={6},
  number={5},
  year={2025},
  publisher={Elsevier},
  url = {https://www.the-innovation.org/article/id/680f5121d3633a2f39d7cb8e}
}

@article{PhysRevB.92.180402,
  title = {Origin of the turn-on temperature behavior in ${\mathrm{WTe}}_{2}$},
  author = {Wang, Y. L. and Thoutam, L. R. and Xiao, Z. L. and Hu, J. and Das, S. and Mao, Z. Q. and Wei, J. and Divan, R. and Luican-Mayer, A. and Crabtree, G. W. and Kwok, W. K.},
  journal = {Physical Review B},
  volume = {92},
  issue = {18},
  pages = {180402},
  numpages = {5},
  year = {2015},
  month = {Nov},
  publisher = {American Physical Society},
  doi = {10.1103/PhysRevB.92.180402},
  url = {https://link.aps.org/doi/10.1103/PhysRevB.92.180402}
}

@article{PhysRevB.96.115152,
  title = {Reentrant metallic behavior in the {Weyl} semimetal {NbP}},
  author = {Xu, J. and Bugaris, D. E. and Xiao, Z. L. and Wang, Y. L. and Chung, D. Y. and Kanatzidis, M. G. and Kwok, W. K.},
  journal = {Physical Review B},
  volume = {96},
  issue = {11},
  pages = {115152},
  numpages = {6},
  year = {2017},
  month = {Sep},
  publisher = {American Physical Society},
  doi = {10.1103/PhysRevB.96.115152},
  url = {https://link.aps.org/doi/10.1103/PhysRevB.96.115152}
}

@article{
doi:10.1073/pnas.1808747115,
author = {Ian A. Leahy  and Yu-Ping Lin  and Peter E. Siegfried  and Andrew C. Treglia  and Justin C. W. Song  and Rahul M. Nandkishore  and Minhyea Lee },
title = {Nonsaturating large magnetoresistance in semimetals},
journal = {Proceedings of the National Academy of Sciences},
volume = {115},
number = {42},
pages = {10570-10575},
year = {2018},
URL = {https://www.pnas.org/doi/abs/10.1073/pnas.1808747115}
}

@article{PhysRevLett.115.046602,
  title = {Temperature-Dependent Three-Dimensional {Anisotropy} of the {Magnetoresistance} in {WTe$_2$}},
  author = {Thoutam, L. R. and Wang, Y. L. and Xiao, Z. L. and Das, S. and Luican-Mayer, A. and Divan, R. and Crabtree, G. W. and Kwok, W. K.},
  journal = {Physical Review Letters},
  volume = {115},
  issue = {4},
  pages = {046602},
  numpages = {5},
  year = {2015},
  month = {Jul},
  publisher = {American Physical Society},
  doi = {10.1103/PhysRevLett.115.046602},
  url = {https://link.aps.org/doi/10.1103/PhysRevLett.115.046602}
}

@article{abrikosov1998quantum,
  title={Quantum magnetoresistance},
  author={Abrikosov, AA},
  journal={Physical Review B},
  volume={58},
  number={5},
  pages={2788},
  year={1998},
  url = {https://doi.org/10.1103/PhysRevB.58.2788},
  publisher={APS}
}

@article{hu2008classical,
  title={Classical and {Quantum} routes to linear magnetoresistance},
  author={Hu, Jingshi and Rosenbaum, TF},
  journal={Nature Materials},
  volume={7},
  number={9},
  pages={697-700},
  year={2008},
  url = {https://www.nature.com/articles/nmat2259},
  publisher={Nature Publishing Group UK London}
}

@article{song2015linear,
  title={Linear magnetoresistance in metals: {Guiding} center diffusion in a smooth random potential},
  author={Song, Justin CW and Refael, Gil and Lee, Patrick A},
  journal={Physical Review B},
  volume={92},
  number={18},
  pages={180204},
  year={2015},
  publisher={APS},
  url = {https://journals.aps.org/prb/abstract/10.1103/PhysRevB.92.180204}
}

@article{lonchakov2024quantum,
  title={Quantum linear magnetoresistance and magnetic-field-induced metal-insulator transition in the {Weyl} semimetal candidate {HgSe}},
  author={Lonchakov, AT and Bobin, SB},
  journal={Journal of Applied Physics},
  volume={135},
  number={23},
  year={2024},
  url = {https://pubs.aip.org/aip/jap/article/135/23/235703/3298677},
  publisher={AIP Publishing}
}

@article{christensen2024extreme,
  title={Extreme magnetoresistance at high-mobility oxide heterointerfaces with dynamic defect tunability},
  author={Christensen, DV and Steegemans, TS and D. Pomar, T and Chen, YZ and Smith, A and Strocov, VN and Kalisky, B and Pryds, N},
  journal={Nature Communications},
  volume={15},
  number={1},
  pages={4249},
  year={2024},
  url = {https://www.nature.com/articles/s41467-024-48398-8},
  publisher={Nature Publishing Group UK London}
}

@article{PhysRevMaterials.7.075002,
  title = {Octahedral distortions in {SrNbO$_3$}: Unraveling the structure-property relation},
  author = {Rosendal, Victor and Brito, Walber H. and Radovic, Milan and Chikina, Alla and Brandbyge, Mads and Pryds, Nini and Petersen, Dirch H.},
  journal = {Physical Review Materials},
  volume = {7},
  issue = {7},
  pages = {075002},
  numpages = {14},
  year = {2023},
  month = {Jul},
  publisher = {American Physical Society},
  doi = {10.1103/PhysRevMaterials.7.075002},
  url = {https://link.aps.org/doi/10.1103/PhysRevMaterials.7.075002}
}

@article{Kresse1996,
  title = {Efficient iterative schemes for ab initio total-energy calculations using a plane-wave basis set},
  author = {Kresse, G. and Furthm\"uller, J.},
  journal = {Physical Review B},
  volume = {54},
  issue = {16},
  pages = {11169-11186},
  numpages = {0},
  year = {1996},
  month = {Oct},
  publisher = {American Physical Society},
  doi = {10.1103/PhysRevB.54.11169},
  url = {https://link.aps.org/doi/10.1103/PhysRevB.54.11169}
}

@article{Pizzi_2020,
doi = {10.1088/1361-648X/ab51ff},
url = {https://doi.org/10.1088/1361-648X/ab51ff},
year = {2020},
month = {jan},
publisher = {IOP Publishing},
volume = {32},
number = {16},
pages = {165902},
author = {Pizzi, Giovanni and Vitale, Valerio and Arita, Ryotaro and Blügel, Stefan and Freimuth, Frank and Géranton, Guillaume and Gibertini, Marco and Gresch, Dominik and Johnson, Charles and Koretsune, Takashi and Ibañez-Azpiroz, Julen and Lee, Hyungjun and Lihm, Jae-Mo and Marchand, Daniel and Marrazzo, Antimo and Mokrousov, Yuriy and Mustafa, Jamal I and Nohara, Yoshiro and Nomura, Yusuke and Paulatto, Lorenzo and Poncé, Samuel and Ponweiser, Thomas and Qiao, Junfeng and Thöle, Florian and Tsirkin, Stepan S and Wierzbowska, Małgorzata and Marzari, Nicola and Vanderbilt, David and Souza, Ivo and Mostofi, Arash A and Yates, Jonathan R},
title = {Wannier90 as a community code: new features and applications},
journal = {Journal of Physics: Condensed Matter},
}

@article{Giannozzi_2009,
doi = {10.1088/0953-8984/21/39/395502},
url = {https://doi.org/10.1088/0953-8984/21/39/395502},
year = {2009},
month = {sep},
publisher = {},
volume = {21},
number = {39},
pages = {395502},
author = {Giannozzi, Paolo and Baroni, Stefano and Bonini, Nicola and Calandra, Matteo and Car, Roberto and Cavazzoni, Carlo and Ceresoli, Davide and Chiarotti, Guido L and Cococcioni, Matteo and Dabo, Ismaila and Dal Corso, Andrea and de Gironcoli, Stefano and Fabris, Stefano and Fratesi, Guido and Gebauer, Ralph and Gerstmann, Uwe and Gougoussis, Christos and Kokalj, Anton and Lazzeri, Michele and Martin-Samos, Layla and Marzari, Nicola and Mauri, Francesco and Mazzarello, Riccardo and Paolini, Stefano and Pasquarello, Alfredo and Paulatto, Lorenzo and Sbraccia, Carlo and Scandolo, Sandro and Sclauzero, Gabriele and Seitsonen, Ari P and Smogunov, Alexander and Umari, Paolo and Wentzcovitch, Renata M},
title = {{QUANTUM ESPRESSO}: a modular and open-source software project for quantum
simulations of materials},
journal = {Journal of Physics: Condensed Matter},
}

@article{PhysRevB.89.214406,
  title = {Anomalous Nernst and {Hall} effects in magnetized platinum and palladium},
  author = {Guo, G. Y. and Niu, Q. and Nagaosa, N.},
  journal = {Physical Review B},
  volume = {89},
  issue = {21},
  pages = {214406},
  numpages = {6},
  year = {2014},
  month = {Jun},
  publisher = {American Physical Society},
  doi = {10.1103/PhysRevB.89.214406},
  url = {https://link.aps.org/doi/10.1103/PhysRevB.89.214406}
}

@article{PhysRevMaterials.4.124203,
  title = {Topological electronic structure and {Weyl} points in nonsymmorphic hexagonal materials},
  author = {Gonz\'alez-Hern\'andez, Rafael and Tuiran, Erick and Uribe, Bernardo},
  journal = {Physical Review Materials},
  volume = {4},
  issue = {12},
  pages = {124203},
  numpages = {20},
  year = {2020},
  month = {Dec},
  publisher = {American Physical Society},
  doi = {10.1103/PhysRevMaterials.4.124203},
  url = {https://link.aps.org/doi/10.1103/PhysRevMaterials.4.124203}
}

@article{PhysRevMaterials.3.124204,
  title = {Symmetry-enforced band crossings in trigonal materials: {Accordion} states and {Weyl} nodal lines},
  author = {Chan, Y.-H. and Kilic, Berkay and Hirschmann, Moritz M. and Chiu, Ching-Kai and Schoop, Leslie M. and Joshi, Darshan G. and Schnyder, Andreas P.},
  journal = {Physical Review Materials},
  volume = {3},
  issue = {12},
  pages = {124204},
  numpages = {12},
  year = {2019},
  month = {Dec},
  publisher = {American Physical Society},
  doi = {10.1103/PhysRevMaterials.3.124204},
  url = {https://link.aps.org/doi/10.1103/PhysRevMaterials.3.124204}
}

@article{Schoop2016,
  author = {L. M. Schoop et al.},
  title = {{Dirac} cone protected by non-symmorphic symmetry and three-dimensional {Dirac} line nodes in {ZrSiS}},
  journal = {Nature Communications},
  volume = {7},
  pages = {11696},
  year = {2016},
  url = {https://www.nature.com/articles/ncomms11696}
}

@Article{Marfoua2022,
author={Marfoua, Brahim
and Hong, Jisang},
title={Reversal of anomalous Hall conductivity by perpendicular electric field in 2D {WSe$_2$/VSe$_2$} heterostructure},
journal={Communications Physics},
year={2022},
month={Oct},
day={30},
volume={5},
number={1},
pages={266},
issn={2399-3650},
doi={10.1038/s42005-022-01044-5},
url={https://doi.org/10.1038/s42005-022-01044-5}
}

@article{Mesbahi2024,
  author  = {Mesbahi, Lyes and Messaoudi, Omar and Bouzar, Hamid and Lounis, Samir},
  title   = {Ab-initio exploration of {Gd} monolayer interfaced with {WSe$_2$}: 
             from electronic and magnetic properties to the anomalous Hall effect},
  journal = {Journal of Physics: Condensed Matter},
  year    = {2024},
  doi     = {10.1088/1361-648X/ae0dd4},
  publisher = {IOP Publishing}
}

@article{supple,
journal={See the Supplemental Material for the additional structural and transport characterisation: 
I. XRD of SL22; II. XRR of SL22 and SL5; III. Temperature-dependent resistivity, MR and Hall resistivity of SL22; IV. Temperature dependent carrier concentration and mobilities of SL5; V. Magnetization(M) vs Field(H), measured at 300 K for SL5; VI. MR along different orientations for SNO; VII. Berry curvature component near Weyl points; VIII. Band structure of the bilayer in the absence of SOC; IX. Projected band structure with the contribution only from the interfacial SNO layer. The crystallographic information files (CIFs) for the relaxed and unrelaxed heterostructures are provided as separate Supporting Information files.
},
url = {https://drive.google.com/file/d/1MHr2wqJTiAOrkTpB5UmA6AKTKr50Yqcm/view?usp=sharing}
}

@article{zhang2017magnetic,
  title={Magnetic two-dimensional electron gas at the manganite-buffered {LaAlO$_3$}/{SrTiO$_3$} interface},
  author={Zhang, HR and Zhang, Y and Zhang, H and Zhang, J and Shen, X and Guan, XX and Chen, YZ and Yu, RC and Pryds, Nini and Chen, YS and others},
  journal={Phys. Rev. B},
  volume={96},
  number={195167},
  pages={10-1103},
  year={2017},
  url = {https://doi.org/10.1103/PhysRevB.96.195167}
}

@article{lazrak2025spontaneous,
  title={Spontaneous nonreciprocal transport in a gate-tunable ferromagnetic {Rashba} 2-dimensional electron gas},
  author={Lazrak, Gabriel and Abrudan, Radu and G{\"o}bel, Borge and Hrabovsky, David and Luo, Chen and Ukleev, Victor and Mallik, Srijani and Vicente-Arche, Luis M and Radu, Florin and Valencia, Sergio and others},
  journal={arXiv preprint arXiv:2508.15424},
  year={2025},
  url = {https://doi.org/10.48550/arXiv.2508.15424}
}

@article{khetan2026stoichiometric,
author = {Khetan, Santosh K. and Bhat, Shwetha G. and Kumar, Vivek and Pradeep, Pranav and Babu, P. D. and Ganguli, Nirmal and Samal, D. and Kumar, P.S. Anil},
title = {Stoichiometric Control of Intrinsic Magnetic Weyl Semimetallic State in SrRuO3 (111) Ultra-Thin Films},
journal = {Advanced Functional Materials},
volume = {n/a},
number = {n/a},
pages = {e30805},
doi = {https://doi.org/10.1002/adfm.202530805},
url = {https://advanced.onlinelibrary.wiley.com/doi/abs/10.1002/adfm.202530805}
}

@article{lin2021electric,
author = {Lin, Weinan and Liu, Liang and Liu, Qing and Li, Lei and Shu, Xinyu and Li, Changjian and Xie, Qidong and Jiang, Peiheng and Zheng, Xuan and Guo, Rui and Lim, Zhishiuh and Zeng, Shengwei and Zhou, Guowei and Wang, Han and Zhou, Jing and Yang, Ping and Ariando and Pennycook, Stephen J. and Xu, Xiaohong and Zhong, Zhicheng and Wang, Zhiming and Chen, Jingsheng},
title = {Electric Field Control of the Magnetic Weyl Fermion in an Epitaxial SrRuO3 (111) Thin Film},
journal = {Advanced Materials},
volume = {33},
number = {36},
pages = {2101316},
doi = {https://doi.org/10.1002/adma.202101316},
url = {https://advanced.onlinelibrary.wiley.com/doi/abs/10.1002/adma.202101316},
year = {2021}
}

@article{jaiswal2022direct,
author = {Jaiswal, Arun Kumar and Wang, Di and Wollersen, Vanessa and Schneider, Rudolf and Tacon, Matthieu Le and Fuchs, Dirk},
title = {Direct Observation of Strong Anomalous Hall Effect and Proximity-Induced Ferromagnetic State in SrIrO3},
journal = {Advanced Materials},
volume = {34},
number = {14},
pages = {2109163},
doi = {https://doi.org/10.1002/adma.202109163},
url = {https://advanced.onlinelibrary.wiley.com/doi/abs/10.1002/adma.202109163},
year = {2022}
}

\end{document}